\documentstyle[epsfig]{mn}
\def\gsimeq{{_>\atop^{\sim}}}
\def\apj{{ApJ.}}
\def\apjs{{ApJS.}}
\def\apjl{{ApJL.}}
\def\nat{{Nature.}}
\def\mnras{{MNRAS}}
\def\araa{{ARA\&A}}
\def\aap{{A\&A}}
\def\aj{{AJ}}

\def\etal{et al.\ }

\def\deg{\hbox{$^\circ$}}
\def\arcmin{\hbox{$^\prime$}}
\def\arcsec{\hbox{$^{\prime\prime}$}}
\def\micron{\hbox{$\mu$m}}

\begin{document}

\title[{\em ELAIS} I:  Goals,  Definition and Observations]
{\vspace{-0.5cm}The European Large Area
{\em ISO} Survey I:  Goals,  Definition and Observations}

\author[Seb Oliver,  \etal]
{\parbox{159mm}{\begin{flushleft}
\vspace{-0.5cm}
Seb Oliver$^{1}$, Michael Rowan-Robinson$^{1}$,\\
{\LARGE D.M.~Alexander$^{2}$,}
{\LARGE O.~Almaini$^{3}$,}
{\LARGE M.~Balcells$^{4}$,}
{\LARGE A.C.~Baker$^{5}$,}
{\LARGE X.~Barcons$^{6}$,}
{\LARGE M.~Barden$^{7}$,}
{\LARGE I.~Bellas-Velidis$^{8}$,}
{\LARGE F.~Cabrera-Guerra$^{4}$,}
{\LARGE R.~Carballo$^{6,9}$,}
{\LARGE C.J.~Cesarsky$^{10}$,}
{\LARGE P.~Ciliegi$^{11}$,}
{\LARGE D.L.~Clements$^{5}$,}
{\LARGE H.~Crockett$^{1}$,}
{\LARGE L.~Danese$^{2}$,}
{\LARGE A.~Dapergolas$^{8}$,}
{\LARGE B.~Drolias$^{1}$,}
{\LARGE N.~Eaton$^{1}$,}
{\LARGE A.~Efstathiou$^{1}$,}
{\LARGE E.~Egami$^{12}$,}
{\LARGE D.~Elbaz$^{10}$,}
{\LARGE D.~Fadda$^{10}$,}
{\LARGE M.~Fox$^{1}$,}
{\LARGE A.~Franceschini$^{13}$,}
{\LARGE R.~Genzel$^{7}$,}
{\LARGE P.~Goldschmidt$^{1}$,}
{\LARGE M.~Graham$^{1}$,}
{\LARGE J.I.~Gonzalez-Serrano$^{6}$,}
{\LARGE E.A.~Gonzalez-Solares$^{4}$,}
{\LARGE G.~L.~Granato$^{13}$,}
{\LARGE C.~Gruppioni$^{11}$,}
{\LARGE U.~Herbstmeier$^{14}$,}
{\LARGE P.~H\'eraudeau$^{14}$,}
{\LARGE M.~Joshi$^{1}$,}
{\LARGE E.~Kontizas$^{8}$,}
{\LARGE M.~Kontizas$^{15}$,}
{\LARGE J.K.~Kotilainen$^{16}$,}
{\LARGE D.~Kunze$^{7}$,}
{\LARGE F.~La~Franca$^{17}$,}
{\LARGE C.~Lari$^{18}$,}
{\LARGE A.~Lawrence$^{3}$,}
{\LARGE D.~Lemke$^{14}$,}
{\LARGE M.J.D.~Linden-V{\o}rnle$^{19,20}$,}
{\LARGE R.G.~Mann$^{1}$,}
{\LARGE I.~M{\'a}rquez$^{21}$,}
{\LARGE J.~Masegosa$^{21}$,}
{\LARGE K.~Mattila$^{22}$,}
{\LARGE R.G.~McMahon$^{23}$,}
{\LARGE G.~Miley$^{24}$,}
{\LARGE V.~Missoulis$^{1}$,}
{\LARGE B.~Mobasher$^{1}$,}
{\LARGE T.~Morel$^{1}$,}
{\LARGE H.~N{\o}rgaard-Nielsen$^{20}$,}
{\LARGE A.~Omont$^{25}$,}
{\LARGE P.~Papadopoulos$^{24}$,}
{\LARGE I.~Perez-Fournon$^{4}$,}
{\LARGE J-L.~Puget$^{26}$,}
{\LARGE D.~Rigopoulou$^{7}$,}
{\LARGE B.~Rocca-Volmerange$^{25}$,}
{\LARGE S.~Serjeant$^{1}$,}
{\LARGE L.~Silva$^{2}$,}
{\LARGE T.~Sumner$^{1}$,}
{\LARGE C.~Surace$^{1}$,}
{\LARGE P.~Vaisanen$^{22}$,}
{\LARGE P.P.~van~der~Werf$^{24}$,}
{\LARGE A.~Verma$^{1}$,}
{\LARGE L.~Vigroux$^{10}$,}
{\LARGE M.~Villar-Martin$^{25}$,}
{\LARGE C.J.~Willott$^{4}$}
\end{flushleft}
}\vspace*{0.200cm}\\  
\parbox{159mm}{
$^{1}$ Astrophysics Group, Blackett Laboratory, Imperial College of 
Science Technology \& Medicine (ICSTM), Prince Consort
Rd., London.SW7 2BZ\\
$^{2}$ SISSA, International School for Advanced Studies, 
Via Beirut 2-4, 34014 Trieste, Italy\\
$^{3}$ Institute for Astronomy, University of Edinburgh, Royal
Observatory, Blackford Hill, Edinburgh EH9 3HJ\\
$^{4}$ Instituto de Astrofisica de Canarias, C/ Via Lactea, s/n,
38200 La Laguna, S/C de Tenerife, Spain\\
$^{5}$ Dept of Physics \& Astronomy, Cardiff University, PO Box 913,
Wales CF24 3YB\\
$^{6}$  Instituto de F\'{\i}sica de Cantabria (Consejo 
Superior de Investigaciones Cient\'{\i}ficas - Universidad de Cantabria), 
39005 Santander, Spain.\\
$^{7}$  Max-Planck-Institut f\"{u}r extraterrestrische Physik,
Postfach 1603, 85740 Garching, Germany\\
$^{8}$ National Observatory of Athens, Astronomical Institute, PO Box
20048, GR-11810, Greece\\
$^{9}$ Departamento de 
F\'{\i}sica Moderna, Universidad de Cantabria, 39005 Santander, Spain\\
$^{10}$ CEA / SACLAY, 91191 Gif sur Yvette cedex, France\\
$^{11}$ Osservatorio Astronomico di Bologna, via Ranzani 1, 
40127 Bologna, Italy\\
$^{12}$ Infrared Astronomy Group, 320-47 Downs Laboratory of Physics, California Institute of
Technology, Pasadena, CA 91125, USA\\
$^{13}$ Dipartimento di Astronomia, Universita' di Padova, Vicolo Osservatorio 5, I-35122 Padova, Italy\\
$^{14}$ Max-Planck-Institut f\"{u}r Astronomie, K\"{o}nigstuhl (MPIA) 17, D-69117, Heidelburg, Germany\\
$^{15}$Section of Astrophysics, Astronomy \& Mechanics, Dept. of Physics, University of Athens, Panepistimiopolis, GR-15783, Zografos, Greece\\
$^{16}$ Tuorla Observatory, University of Turku, V\"ais\"al\"antie 20, FIN-21500 Piikki\"o, Finland\\
$^{17}$ Dipartimento di Fisica, Universita degli Studi ``Roma TRE''
Via della Vasca Navale 84, I-00146, Roma, Italy\\
$^{18}$ Institute di Radio Astronomy, Bologna, Italy\\
$^{19}$Niels Bohr Institute for Astronomy, Physics and Geophysics,
Astronomical Observatory, Juliane Maries Vej 30, DK--2100 Copenhagen
{\O}, Denmark\\
$^{20}$ Danish Space Research Institute, Juliane Maries Vej 30,
DK--2100 Copenhagen {\O}, Denmark\\
$^{21}$  Instituto de Astrof\'{i}sica de Andaluc\'{i}a, CSIC, Apartado 3004,
 E-18080 Granada, Spain.\\
$^{22}$Observatory, P.O. Box 14, Tahtitorninmaki, FIN-00014 University
of Helsinki, Finland\\
$^{23}$ Institute of Astronomy, The Observatories, Madingley Road, Cambridge, CB3 0HA\\ 
$^{24}$ Leiden Observatory, P.O. Box 9513,  NL-2300 RA Leiden, The
Netherlands\\ 
$^{25}$ Institut d'Astrophysique de Paris, 98bis Boulevard Arago, F 75014 Paris, France\\ 
$^{26}$ Institut d'Astrophysique Spatiale (IAS),  B\^{a}timent 121, Universit\'{e} Paris XI, 91405 Orsay cedex, France\\
}}

%
%
\date{Accepted ;
      Received ;
      in original form 11 June 1999}

\pagerange{\pageref{firstpage}--\pageref{lastpage}}
\pubyear{1999}
\volume{999}

\label{firstpage}

\maketitle


\begin{abstract}
We describe the European Large Area {\em ISO\/} Survey (ELAIS).
ELAIS was the largest single Open Time project conducted by {\em
ISO\/}, mapping an area of 12 square degrees at 15$\mu$m with 
{\em ISO-CAM\/} and at 90$\mu$ with {\em ISO-PHOT\/}. Secondary
surveys in other {\em ISO\/} bands were undertaken by the ELAIS team 
within the fields of the primary survey, with  6 square degrees being
covered at 6.7$\mu$m and 1 square degree at 175$\mu$m. 

This paper discusses the goals of the project and the techniques employed
in its construction, as well as presenting details of the observations carried
out, the data from which are now in the public domain.  
We outline the ELAIS ``Preliminary Analysis'' which led to the detection of
over 1000 sources from the 15 and 90 $\mu$m surveys (the majority
selected at 15$\mu$m with a flux limit
of $\sim$3 mJy), to be fed
into a ground--based follow--up campaign, as well as
a programme of photometric observations of detected sources using both
{\em ISO-CAM\/} and {\em ISO-PHOT\/}. 

We detail how the ELAIS survey complements other {\em ISO\/} surveys in
terms of depth and areal coverage, and show that the extensive  
multi--wavelength coverage of the ELAIS fields resulting from our
concerted and on--going follow--up programme has made these regions amongst 
the best studied areas of their size in the entire sky, and, therefore, natural
targets for future surveys.
This paper accompanies the release of extremely reliable 
sub-sets of the ``Preliminary Analysis'' products.
Subsequent papers in this series will give further details of our data
reduction techniques, reliability \& completeness estimates and 
present the 15 and 90 $\mu$m number
counts from the ``Preliminary Analysis'', while a further series of
papers will discuss in detail the results from the ELAIS ``Final
Analysis'', as well as from the follow--up programme.

\end{abstract}



\section{Introduction}

The {\em Infrared Space Observatory} 
\cite{Kessler et al. 1996} was the
natural successor to the {\em Infrared Astronomical Satellite} ({\em IRAS}), and has primarily been used to undertake
detailed studies of individual objects and regions.  However, {\em ISO}
also provided an opportunity to perform survey work at sensitivities
beyond the reach of {\em IRAS}.  The {\em IRAS} survey was of
profound significance for cosmology, extragalactic astrophysics and
for the study of stars, star-forming regions and the interstellar
medium in the Galaxy.  The mapping of large-scale structure
\cite{Saunders et al. 1991} in the
galaxy distribution, the discovery of ultra-luminous infrared galaxies
(see the review by Sanders \& Mirabel \shortcite{Sanders et al. 1996})
and of hyper-luminous infrared galaxies like {\em IRAS} F10214+4724
\cite{Rowan-Robinson et al. 1991a}, and the detection of proto-planetary
discs around fairly evolved stars, were all  unexpected
discoveries of the {\em IRAS} survey.  The $z = 2.3$ galaxy
F10214+4724, was at the limit of detectability by {\em IRAS}
($S_{60}\gsimeq 0.2$Jy).  Several
other $z > 1$ galaxies and quasars have now been found from follow-up
of faint {\em IRAS} samples.  Recent sub-mm surveys,  in particular with SCUBA
on the JCMT, (e.g. 
Smail et al. \shortcite{Smail et al. 1997}
Hughes et al.\shortcite{Hughes et al. 1998} 
Barger et al. \shortcite{Barger et al. 1998} 
Eales et al.  \shortcite{Eales et al. 1999} 
Blain et al.  \shortcite{Blain et al. 1999}  
) are
detecting sources which are probably very high redshift
counterparts to these {\em IRAS} sources.  Pointed observations of
high redshift quasars and radio galaxies produce detections at
sub-millimetre wavelengths in continuum and line emission, but mostly
lie below the limit of the {\em IRAS} survey at far infrared
wavelengths.

While designed as an observatory instrument, the  huge
improvement in sensitivity provided by  {\em ISO}   offered the opportunity to probe
the galaxy population to  higher redshift than {\em IRAS} and to
make progress in understanding the obscured star formation history 
of the Universe.  A significant fraction of the mission time was thus
spent on field surveys.
In this paper we describe the
``European Large Area {\em ISO} Survey'' ({\em ELAIS}) 
which represents the largest non-serendipitous survey 
   conducted with {\em ISO}.  This survey
provides a link between the {\em IRAS} survey, the deeper {\em ISO}
surveys and the sub-mm surveys.  

{\em ELAIS} is a collaboration involving  25 European
institutes, led from Imperial College.
This project surveyed around 12  square degrees of the sky at
15$\mu$m and 90$\mu$m nearly  6 square degrees at 6.7$\mu$m together with a
further one square degree at 175$\mu$m.  The survey
used the {\em ISO} Camera \cite{Cesarsky et al. 1996} at the 
two shorter wavelengths and the {\em ISO} Photometer  \cite{Lemke et al. 1996} at the longer wavelengths.
{\em ELAIS} was the largest open time project undertaken by {\em ISO}: a total
of 375 hours of scientifically validated data have been produced.
We have detected over 1000 extra-galactic objects and a
similar number of Galactic sources.  Around 200 of these objects have
been re-observed with {\em ISO} to provide detailed mid/far infrared
photometry.

This paper outlines the broad scientific objectives of this project
and describes the selection of the observing modes and survey fields.
It also details the execution of the {\em ISO} observations and briefly
outlines the data reduction and data products. Finally we show how
this survey complements other {\em ISO} surveys and summarise the
extensive multi-wavelength programmes taking place in the {\em ELAIS}  fields.

\section{Key Scientific Goals}

\subsection{The Star Formation History of the Universe}

The main extra-galactic population detected by {\em IRAS} was galaxies
with high rates of star formation. These objects are now known to
evolve with a strength comparable to Active Galactic Nuclei (AGN) (e.g. Oliver et al. \shortcite{Oliver et al. 1995}).
The distance to which these objects were visible to {\em IRAS} was,
however, insufficient to determine the nature of their evolution.  The
sensitivity of {\em ISO} allows us to detect these objects at much
higher redshifts and thus obtain greater understanding of the
cosmological history of star formation.  The infrared luminosity
provides a better estimate of the total star formation rate than
optical and UV estimators (e.g. Madau et al.\shortcite{Madau et al. 1996}) as these monitor
star formation only from regions with low obscuration and require
large corrections for extinction \cite{Steidel et al. 1999}.  Another
important star formation indicator for galaxies is the radio
luminosity (e.g. Condon \shortcite{Condon 1992}). For galaxies obeying the well known
far infrared radio correlation \cite{Helou et al. 1985}, the depth of the
survey described here is well matched to that of sub-mJy radio surveys
(e.g. 
Condon \& Mitchell \shortcite{Condon et al. 1984},           
Windhorst      \shortcite{Windhorst 1984},                    
Windhorst et al. \shortcite{Windhorst et al. 1985},             
Hopkins et al.   \shortcite{Hopkins et al. 1998},               
Gruppioni, Mignoli \& Zamorani \shortcite{Gruppioni et al. 1999b}).  
Comparison of the global star formation rate
determined in the infrared with other determinations from the optical
and UV luminosity densities, $H\alpha$ luminosity density, radio
luminosity density, etc. will give a direct estimate of the importance
of dust obscuration, vitally important for models of cosmic evolution,
as well as providing us with a reliable estimate for the total
star formation rate.  The {\em ELAIS} follow-up surveys (see section
\ref{followup}) will allow us to go a stage further and apply a number
of these complementary star formation tracers to the same volume and
in many cases on the same objects, thereby addressing the impact of
dust extinction independently of any peculiarities to any particular
survey volume.

\begin{figure}
\epsfig{file=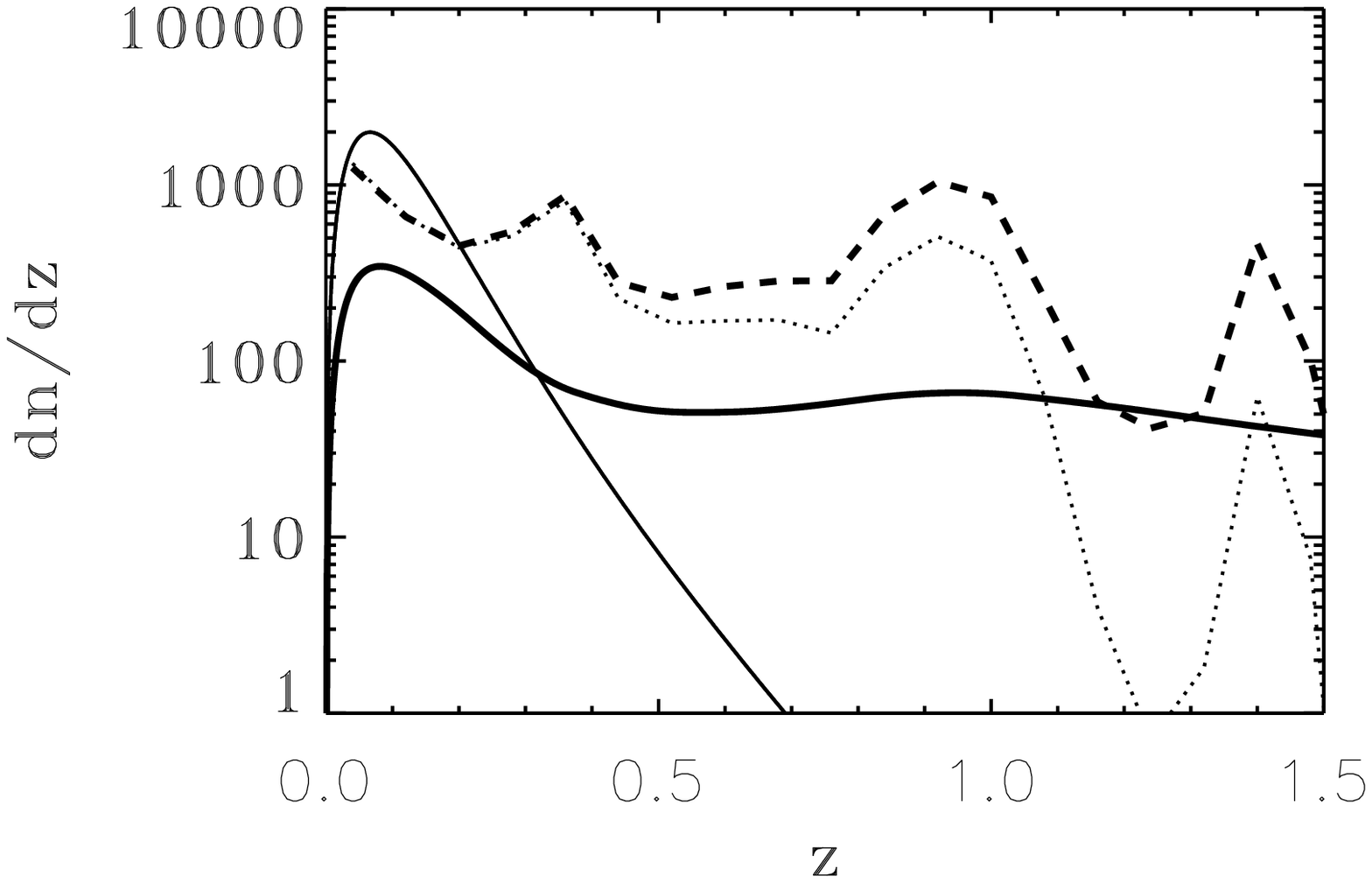, width=8cm}
\caption{Expected populations and redshift distribution for 
the 15\micron\ survey,
assuming a depth of 3mJy over 11 square degrees.  The redshift
distribution is plotted as $dn/dz$ in dimensionless units.
Cirrus (solid line) and Star-burst (thick solid line) components are from a model
similar to Oliver et al. (1992) and Pearson \& Rowan-Robinson (1996),
model {\em E}  (dashed line) and 
model {\em A}  (dotted line) are from Guiderdoni et al. (1998).
}\label{fig:nofz15}
\end{figure}

\begin{figure}
\epsfig{file=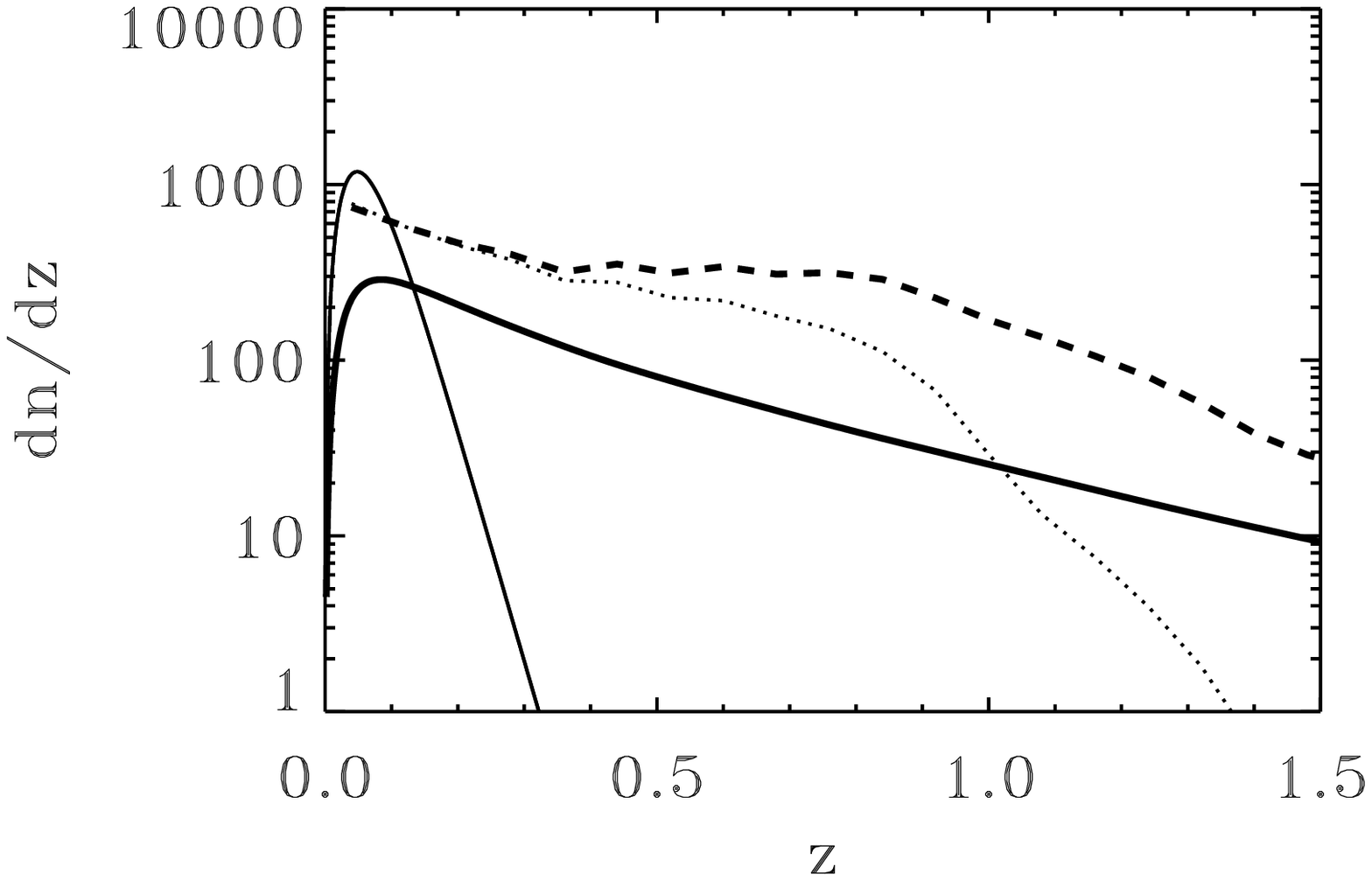, width=8cm}
\caption{Expected populations and redshift distribution for 
the 90\micron\ survey,
assuming a depth of 100mJy over 11 square degrees. The redshift
distribution is plotted as $dn/dz$ in dimensionless units.
Cirrus (solid line) and Star-burst (thick solid line) components are from a model
similar to Oliver et al. (1992) \shortcite{Oliver et al. 1992} a and Pearson \& Rowan-Robinson (1996),
model {\em E}  (dashed line) and 
model {\em A}  (dotted line) are from Guiderdoni et al. (1998).
}\label{fig:nofz90}
\end{figure}

\begin{figure}
\epsfig{file=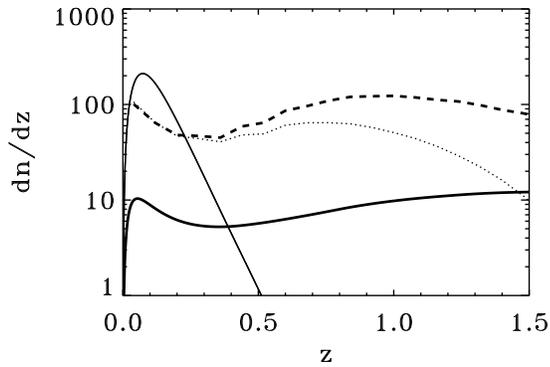, width=8cm}
\caption{Expected populations and redshift distribution for 
the 175\micron\ survey,
assuming a depth of 75mJy over 1 square degrees. The redshift
distribution is plotted as $dn/dz$ in dimensionless units.
Cirrus (solid line) and Star-burst (thick solid line) components are from a model
similar to Oliver et al. (1992) and Pearson \& Rowan-Robinson (1996),
model {\em E}  (dashed line) and 
model {\em A}  (dotted line) are from Guiderdoni et al. (1998).
}\label{fig:nofz175}
\end{figure}

Figures \ref{fig:nofz15}-\ref{fig:nofz175},
show the predicted redshift distribution of star-forming galaxies
in the {\em ELAIS} survey selected at 15\micron, 90\micron\ and
175\micron. The predictions come 
from three different evolutionary models; the
first model is that of Pearson \& Rowan-Robinson 
\shortcite{Pearson et al. 1996},
 the second and third are models `A' and `E' from Guiderdoni
et al. \shortcite{Guiderdoni et al. 1998}. All three models are extrapolations from {\em IRAS} data.
The total number of objects of various different types
predicted by two of these models and a
third from Franceschini et al. \shortcite{Franceschini et al. 1994}, are also tabulated in Table
\ref{tab:nsource}.  While the source counts from {\em ELAIS}
alone may not be able to distinguish between such models,
spectroscopic identifications, source classifications and the redshift
distributions will.

\begin{table*}

\begin{tabular}{lccc}
Model   & 
Pearson \& Rowan-Robinson \shortcite{Pearson et al. 1996}
& Franceschini et al. \shortcite{Franceschini et al. 1994}
& Guiderdoni et al. \shortcite{Guiderdoni et al. 1998}\\
\\
&\multicolumn{3}{c} {6.7\micron} \\
Elliptical            &  &  8 & \\
Normal spiral         & 455 & 38 & \\
Star-forming galaxies & 122 &   & \\
AGN                   &  64 & 31 & \\

\\
&\multicolumn{3}{c} {15\micron}\\
Elliptical            &   &  11 & \\
Normal spiral         & 308  & 378 &    \\
Star-forming galaxies &  181 & 177 & 258\\
AGN                   &  112 &  14 &    \\
\\
&\multicolumn{3}{c} {90\micron}\\
Elliptical            &      &    &    \\
Normal spiral         & 106  &    &    \\
Star-forming galaxies & 109  &    & 231\\
AGN                   &      &    &    \\
\\
&\multicolumn{3}{c} {175\micron} \\
Elliptical            &     &     &    \\
Normal spiral         & 102 &     &    \\
Star-forming galaxies &  76 &     & 261\\
AGN                   &     &     &    \\
\\

\end{tabular}
\caption{Expected numbers of extra-galactic sources from three a priori
evolutionary models. Assuming survey areas of 6, 11, 11 \& 3 
and depths of 1, 3, 100, 75 mJy in the
6.7, 15, 90 \&  175\micron bands respectively (roughly
those achieved in our Preliminary Analysis). The Pearson \&
Rowan-Robinson (1996) models do not include Elliptical populations, the
predictions at 175\micron\ are in fact calculated at 200\micron. The
Franceschini et al. (1994) models were not available for longer wavelengths,
the Guiderdoni et al. model (1998) is their strongly evolving model `E',
while their models include objects of different luminosities and SEDs
their results do not discriminate so we arbitrarily assign them all to the
star-forming row.}\label{tab:nsource}
\end{table*}

\subsection{Ultra-luminous Infrared Galaxies at High $z$}\label{ultlum}

{\em IRAS} uncovered a population with enormous far infrared luminosities,
$L_{\rm FIR} $ $> 10^{12}L_{\odot}$ (see the review by Sanders \& Mirabel 1996).    While somewhere between 20 and
50 per cent of
these objects appear to have an AGN 
(Veilleux et al. \shortcite{Veilleux et al. 1995},
Sanders  et al. \shortcite{Sanders et al. 1999},
Veilleux et al. \shortcite{Veilleux et al. 1999},
Lawrence et al. \shortcite{Lawrence et al. 1999}) it is still a source of
controversy as to whether the  illumination  of the dust
arises principally from an AGN or a star-burst.
 {\em ISO} spectra of samples of ultra-luminous infrared galaxies
(Genzel et al.      \shortcite{Genzel et al. 1998},
Lutz et al.      	\shortcite{Lutz et al. 1998},     
Lutz, Veilleux \& Genzel\shortcite{Lutz et al. 1999}, 
Rigopoulou et al. 	\shortcite{Rigopoulou et al. 1999}) 
appear to demonstrate
that  while some do require the photoionization energies typical for
AGN to explain the obscured lines, most  are consistent with star-burst
models.
Interestingly, most of these
objects appear to be in interacting systems, suggesting a
 mechanism that could trigger either an AGN, a star-burst, or indeed
both (e.g., 
Sanders et al.  \shortcite{Sanders et al. 1988}, 
Lawrence et al.  \shortcite{Lawrence et al. 1989}, 
Leech et al.  \shortcite{Leech et al. 1994},
Clements et al.  \shortcite{Clements et al. 1996}).

\begin{figure}
\epsfig{file=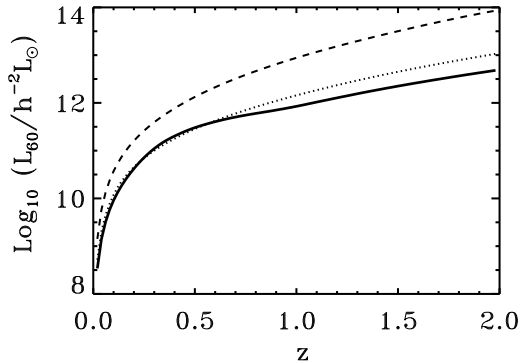,width=8cm}
\caption{The minimum rest-frame 60\micron\ luminosity for a source
with a star-burst spectrum detectable in the {\em ELAIS} survey at 
$S_{15}>3$mJy (solid line), $S_{90}>100$mJy (dotted line), 
$S_{175}>75$mJy (dot-dashed line) and in the {\em IRAS}
survey $S_{60}>250$mJy.(dashed line)}\label{fig:lumz}
\end{figure}

The area of this survey is small compared to that of {\em IRAS} so we
would not expect to detect large numbers of these objects.  
The Pearson \& Rowan-Robinson (1996) model would predict that we would detect
between 40 and 80  of these objects, though models such as that of 
Guiderdoni et al. (1998), which takes into account the increase in temperature
of the dust with  increasing luminosity
would predict more.
Nevertheless such objects will be
visible at greater distances than they were in {\rm IRAS} and even a
few examples at higher redshift would be interesting. Assuming a 
star-burst SED \cite{Rowan-Robinson et al. 1993} an object of
$L_{60}=10^{12}L_{\odot}$ ($H_0=50, q_0=1/2$) would be visible
($S_{15}>3{\rm mJy}$) in the
{\em ELAIS} survey to 
$z=0.5$ where it is only visible to $z=0.26$ in the {\em IRAS Faint
Source Catalog} ($S_{60}>0.2{\rm Jy}$) and to $z=0.15$ in the
{\em IRAS Point
Source Catalog} ($S_{60}>0.6{\rm Jy}$).
{\em ELAIS} thus allows us to study samples of these controversial
objects at higher redshift where both AGN and star formation 
are known to be enhanced.  Figure \ref{fig:lumz} shows the 
minimum 60\micron\ lluminosity of a source which could be detected in
both the {\em ELAIS} survey and the {\em IRAS} survey as a function of redshift. 

\subsection{Emission from Dusty Tori around AGN}

The orientation-based unified models of AGN involve a central engine
surrounded by an optically and geometrically thick torus 
(Antonucci \& Miller \shortcite{Antonucci et al. 1985},
Scheuer  		   \shortcite{Scheuer 1987}, 
Barthel  		   \shortcite{Barthel 1989}, 
Antonucci 		   \shortcite{Antonucci 1993}). In this model
the optical properties of the central regions are dependent on the
inclination angle of the torus, with type 2 objects defined as those with
the central nucleus obscured by the torus, and type 1 objects (such as
quasars) as those with an unobscured view of the nucleus. Objects with
radio jets have the jets aligned approximately with the torus symmetry
axis. The scheme is very attractive in providing a single conceptual
framework for what would otherwise appear to be extremely diverse
populations, and the models have survived many observational tests and
predictions. It is now widely accepted that the unified models are broadly
correct at least to ``first order'' (e.g. Antonucci 1993) and that many if
not most type 2 AGN contain obscured type 1  nuclei.

An important corollary of the unified models is the expectation that  populations of obscured
(i.e. type 2) AGN will be present all redshifts. These predicted populations are in
general extremely difficult to identify observationally 
(e.g. Halpern \& Moran \shortcite{Halpern et al. 1998}) except locally in low-luminosity AGN, and at high redshift
($0<z\stackrel{<}{_\sim}5$) in the radio-loud AGN minority. Nevertheless,
the strength and shape of the X-ray background has been taken as evidence
of the existence of a large population of obscured quasars, outnumbering
normal quasars by a factor of several (e.g. Comastri et al. \shortcite{Comastri et al. 1995}). Such a
large population of obscured quasars may also explain the unexpectedly
large population of local remnant black holes 
(Fabian \& Iwasawa \shortcite{Fabian et al. 1999},
Lawrence \shortcite{Lawrence 1999}). Even hard X-ray samples may miss the very heavily obscured
objects, so an infrared-selected sample is the only reliable way to obtain
a complete census of AGN. For example, it will be possible with {\em ELAIS} to
make quantitative constraints on the dust distribution and torus column
densities, as well as on the evolution of obscured quasar activity.

\subsection{Dust in Normal Galaxies to Cosmological Distances}

At the longer {\em ISO} wavelengths (90 and 175 \micron) emission from
the cool  interstellar `cirrus' dust
in normal galaxies will be detectable in our survey 
in fainter and cooler objects than were accessible to {\em
IRAS}. This will allow us to
examine the temperature distribution functions and in particular look
for unusually cool galaxies.
Quantifying the distributions of such cool sources will be important
for deep sub-mm surveys as there is considerable degeneracy between
cool,  low redshift and warm high redshift objects in this
wavelength regime.

\subsection{Circumstellar Dust Emission from Galactic Halo Stars}

We expect to detect hundreds of stars at 6.7 and 15 microns and it will be
of interest to check whether any show evidence of an infrared excess due
to the presence of a circumstellar dust shell.  Such shells are expected
from late type stars due to mass-loss while on the red giant branch, from
cometary clouds or from proto-planetary discs. At the high
galactic latitudes of our survey, late type stars with circumstellar
dust shells  should be rare (e.g. Rowan-Robinson \& Harris,
\shortcite{Rowan-Robinson et al. 1983}), so any
detections of such shells could be especially interesting.

\subsection{New classes of Galactic and Extra-Galactic Objects}

F10214+4724  \cite{Rowan-Robinson et al. 1991a}
was at the limit of {\em IRAS} sensitivity  and new classes of objects
may well be discovered at the limit of the {\em ELAIS} sensitivity.  The
lensing phenomenon which made F10214+4724 detectable by {\em IRAS} may become more
prevalent at fainter fluxes, increasing the proportion of interesting
objects.

\subsection{The  Extra-Galactic Background}\label{firback}

The discovery of the $140-850$\micron\ far infrared background 
(Puget et al. \shortcite{Puget et al. 1996}, 
Fixsen et al. \shortcite{Fixsen et al. 1998}, 
Hauser et al. \shortcite{Hauser et al. 1998}, 
Lagache et al \shortcite{Lagache et al. 1998}) 
from {\em COBE} data
has shown that most of the light produced by extra-galactic objects
has been reprocessed by dust and re-emitted in the far infrared and
sub-mm.  This discovery provides further strong  motivation for
studying the dust emission from objects at all redshifts and all
far infrared wavelengths.  It is possible to explain this far infrared
background radiation with a number of evolution models that are
consistent with the {\em IRAS} data.
The constraints provided by {\em ISO} surveys such as {\em ELAIS} are
expected to be able to rule out some of these a priori models.  
The motivation behind our 175\micron\ survey was specifically 
to start to resolve this far infrared background into its constituent
galaxies.

\section{Survey Definition}

\subsection{Selection of survey wavelengths and area}\label{survey_lam}

In order to detect as many sources as efficiently as possible we
restricted ourselves to two primary {\em ISO} broad band filters and
aimed to cover as large an area as possible.  We selected filters with
central wavelengths at: 15\micron\ \cite{Cesarsky et al. 1996}
 which is particularly sensitive to AGN emission and
90\micron\ \cite{Lemke et al. 1996} which is sensitive
to emission from star formation regions.  At 90\micron\ we aimed to
reach the confusion limit and pre-flight sensitivity estimates led us
to conclude that this could be achieved with an on-sky integration
time of 20s.  We decided to map the same area of sky at 15\micron\
using a similar total observation time and this required on-sky
integration times of 40s.  In both cases these integration times were
close to the minimum practical.  A survey area of order 10 square
degrees was chosen to produce a statistically meaningful sample of
galaxies.  This area and depth was ideal to complement the deep {\em
ISO-CAM} surveys (Cesarsky et al. (1996), Elbaz et al. \shortcite{Elbaz et al. 1999}, 
Taniguchi et al. \shortcite{Taniguchi et al. 1997a}) as discussed in Section \ref{isosurveys}.

\begin{figure}
\epsfig{file=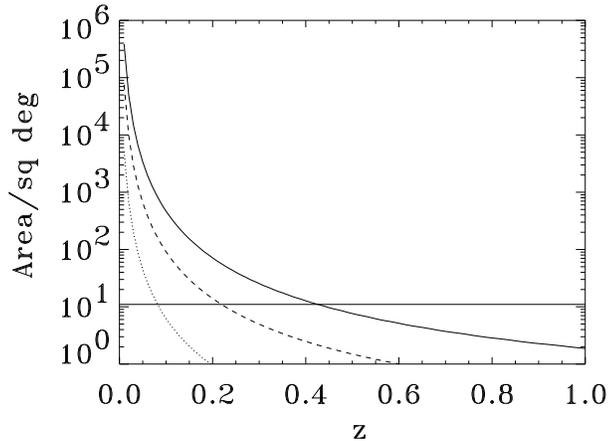, width=6cm, angle=90}

\caption{The minimum area of a survey required to measure mean densities in
populations visible to a given redshift such that the systematic
errors due to large-scale structure are: $\sigma=0.1$ - solid line,
$\sigma=0.2$ - dashed line, $\sigma=0.5$ - dotted line.   The nominal
area of the {\em ELAIS} survey is over-plotted.  This plot assumes
that the survey area is split into four independent survey areas
as is the case for {\em ELAIS}. 
}\label{fig:cosmic_var} 
\end{figure}

\begin{figure}
\epsfig{file=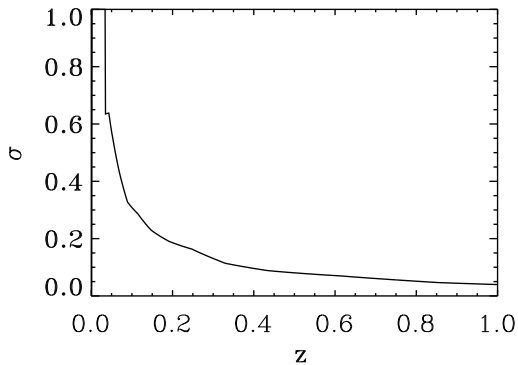, width=8cm}

\caption{The fractional error in the  mean densities due to cosmic
variance for
populations visible to a given redshift for the {\em ELAIS} survey.
}\label{fig:cosmic_var2} 
\end{figure}

A further justification for a large area survey is that many of the
sources will be at relatively low redshift (e.g. an ultra-luminous
star-burst would be detectable at $z=0.5$ as discussed in Section
\ref{ultlum}).  Thus, unless our survey is of a sufficient area, the
volume will be such that cosmic variance can be a significant problem,
i.e. large-scale clustering means that the mean density within a
survey volume may not be representative of the universal mean.  To
estimate this effect we use the galaxy power spectrum as compiled by
Peacock \& Dodds \shortcite{Peacock et al. 1994}. From this we can estimate the variance in a
survey of any given volume (we assume a cubical geometry, which means
we will underestimate the variance).  Figure \ref{fig:cosmic_var}
illustrates the area required to study populations out to a given
redshift allowing for different amounts of cosmic variance. From this
we can see that a survey of around 10 square degrees is required to
measure the mean density of populations visible to $z=0.5$ with
negligible errors ($<10$ per cent) due to large-scale structure.  A
survey with the area of {\em ELAIS} can also measure the mean density
of populations $z=0.25$ with 20 per cent accuracy.  Populations below
$z=0.15$ would only have mean densities known to around 50 per cent.
Figure \ref{fig:cosmic_var2} shows what fractional errors we would expect 
in mean quantities derived from {\em ELAIS} for populations that are 
visible to different depths.

During the mission we introduced two additional filters.  The first of
these was designed to provide constraints on the infrared spectral
energy distribution of {\em ELAIS} sources from fields (around six
square degrees) that would not have been observed in time for pointed
{\em ISO} follow-up.  For  this aspect of the
survey we selected the 6.7\micron\ filter, which was the most sensitive
for sources detected at 15$\mu$m. 
Naturally as well as providing improved spectral coverage
of other {\em ELAIS} sources this also produced an independent source
list which was sensitive to emission from normal galaxies.
The second filter, centred at 175\micron, was introduced
specifically to explore the populations making up the far infrared
background as discussed in Section \ref{firback}.

A more detailed description of the survey parameters is given Section \ref{obsparam}.

\subsection{Time Awarded}\label{times}

Over the course of the {\em ISO} mission the {\em ELAIS} programme was
awarded a total of 377 hours.  This allocation was used not only to
perform the basic blank field survey observations, discussed in
Section~\ref{survey_lam}, but also a number of other related
programmes.

Principal among these was an {\em ISO} photometry programme to
investigate around 200 sources that had been detected by {\em ELAIS}
in the early parts of the mission.  These observations were designed
to provide constraints on the spectral energy distributions of the
{\em ELAIS} sources but would also provide a serendipitous, though
biased, survey in
their own right.  In addition we were awarded time to observe a number
of sub-fields repeatedly to help quantify our reliability and
completeness.   We also performed eight {\em ISO-PHOT} calibration
measurements on three known stars and three {\em ELAIS} sources,
independently of the instrument team.

The amount of time actually spent and AOTs used on both the survey proper and
the photometry programmes are summarised in Table~\ref{tab:times}.

\begin{table}
\begin{tabular}{lrr}
Category               & Time/hr & Number of AOTS \\
                       &         &                \\
Awarded                & 377     &                \\
                       &         &                \\
Survey (Observed)      & 324     & 174            \\
Survey (Aborted)       &   7     &   3            \\
Survey (Failed)        &   5     &   3            \\
                       &         &                \\
Photometry (Observed)  &  44     & 930            \\

Total (Observed)            &  368    & 1104           \\
Total (Observed \& Aborted) &  375    & 1107           \\

\end{tabular}
\caption{Summary of the total time spent by {\em ISO} on the {\em ELAIS}
programme.  This table is broken down into observations preformed for
the survey (including reliability observations, calibration
measurements and small fields) and the photometry section of the programme.  Data
flagged as ``aborted'' or ``failed'' by the {\em ISO} ground station are
also singled out,  however we have not experienced any problems with
the ``aborted'' data.}\label{tab:times}
\end{table}

\subsection{Field Selection}\label{field_selection}

\begin{figure*}
\epsfig{file=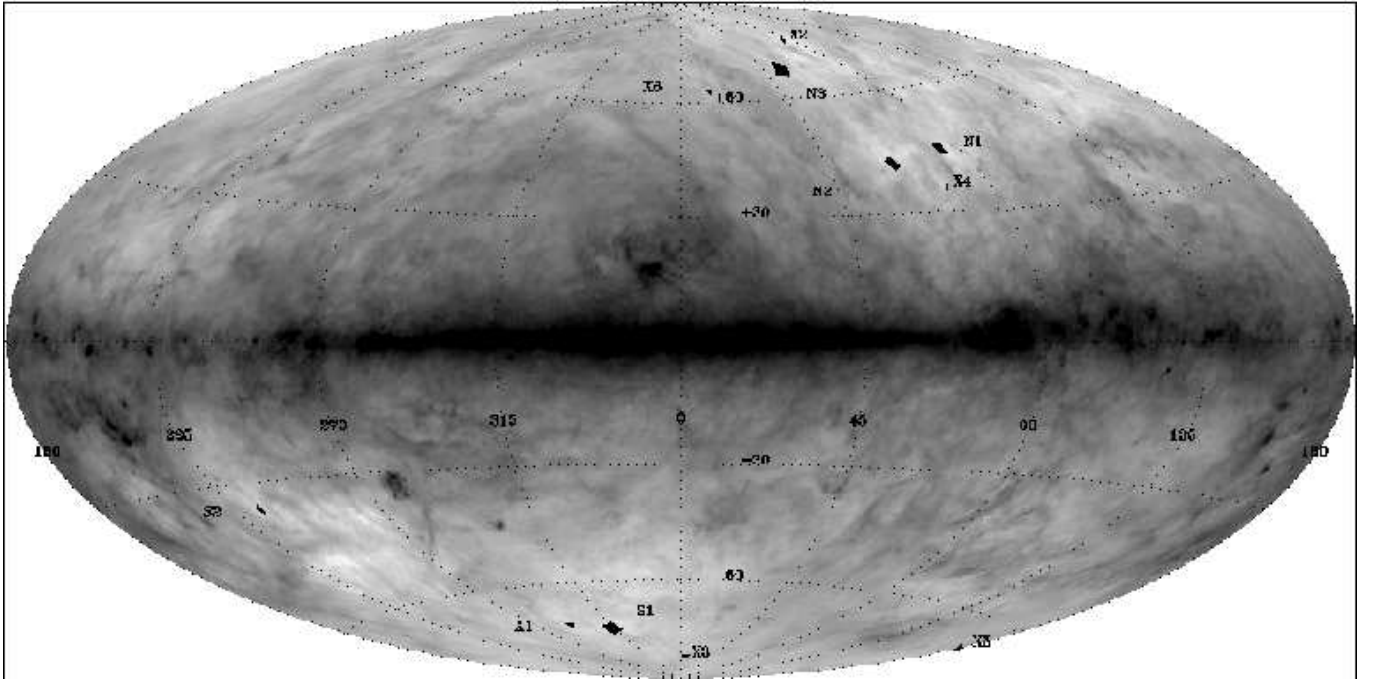,width=18cm}

\caption{The location of the  {\em ELAIS} survey
fields overlaid on a Hammer-Aitoff equal area projection of the {\em COBE} normalised {\em IRAS} maps of
Schlegel et al. (1998).   Galactic latitude and longitude gridlines are
overlaid.
}\label{fig:i100allsky}
\end{figure*}

\begin{figure}
\epsfig{file=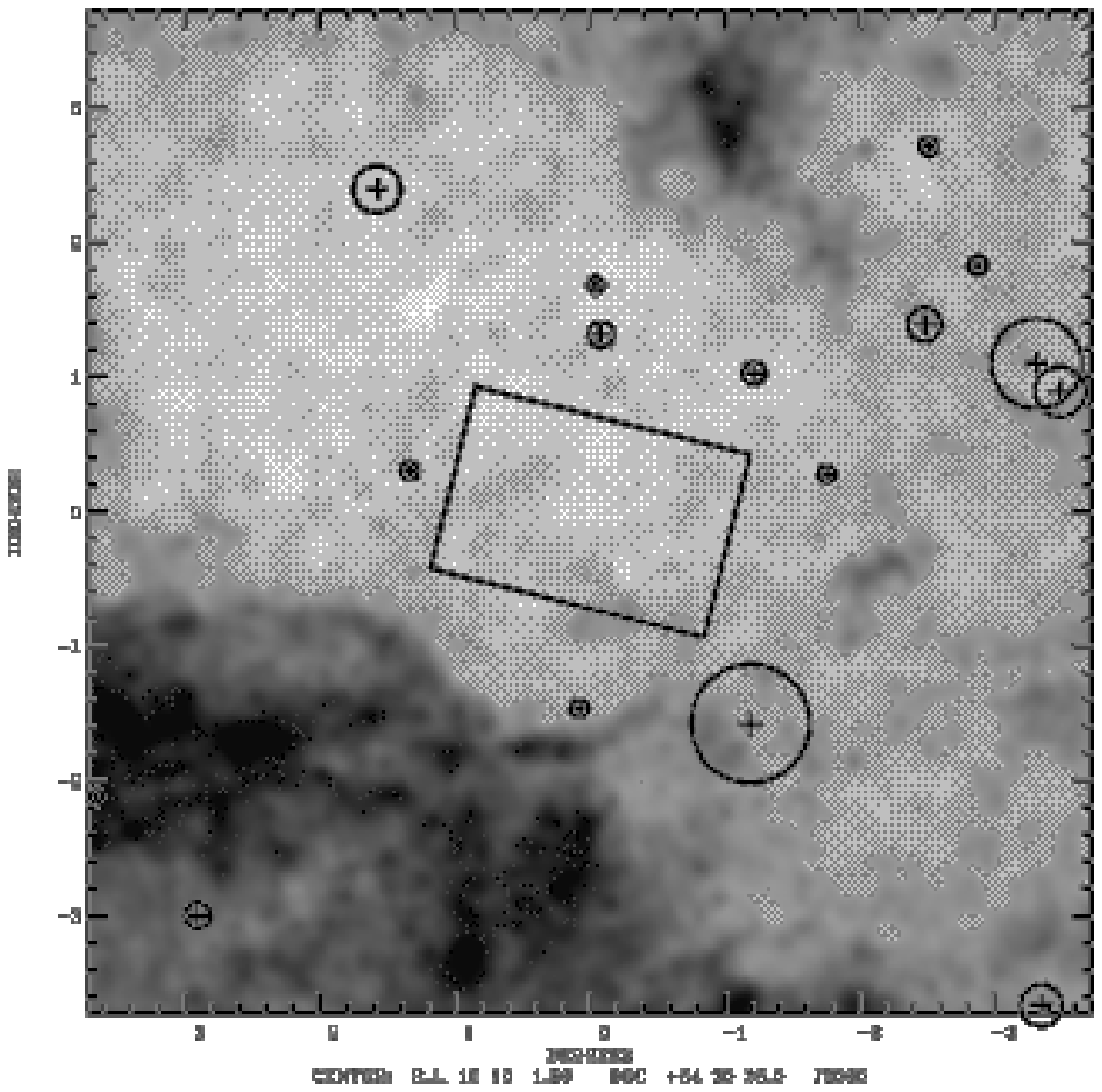,width=8cm}
\caption{The location of the N1 {\em ELAIS} survey
field overlaid on the {\em COBE} normalised {\em IRAS} maps of 
Schlegel et al. (1998).  {\em IRAS} sources with 12\micron\ fluxes brighter than 0.6Jy are
overlayed with radius proportional to flux for $S_{60}<10$Jy.  
The minimum 100\micron\ intensity shown is 0MJy/sr (white)
and the maximum is 1.5MJy/sr (Black).}\label{fig:i100n1}
\end{figure}

\begin{figure}
\epsfig{file=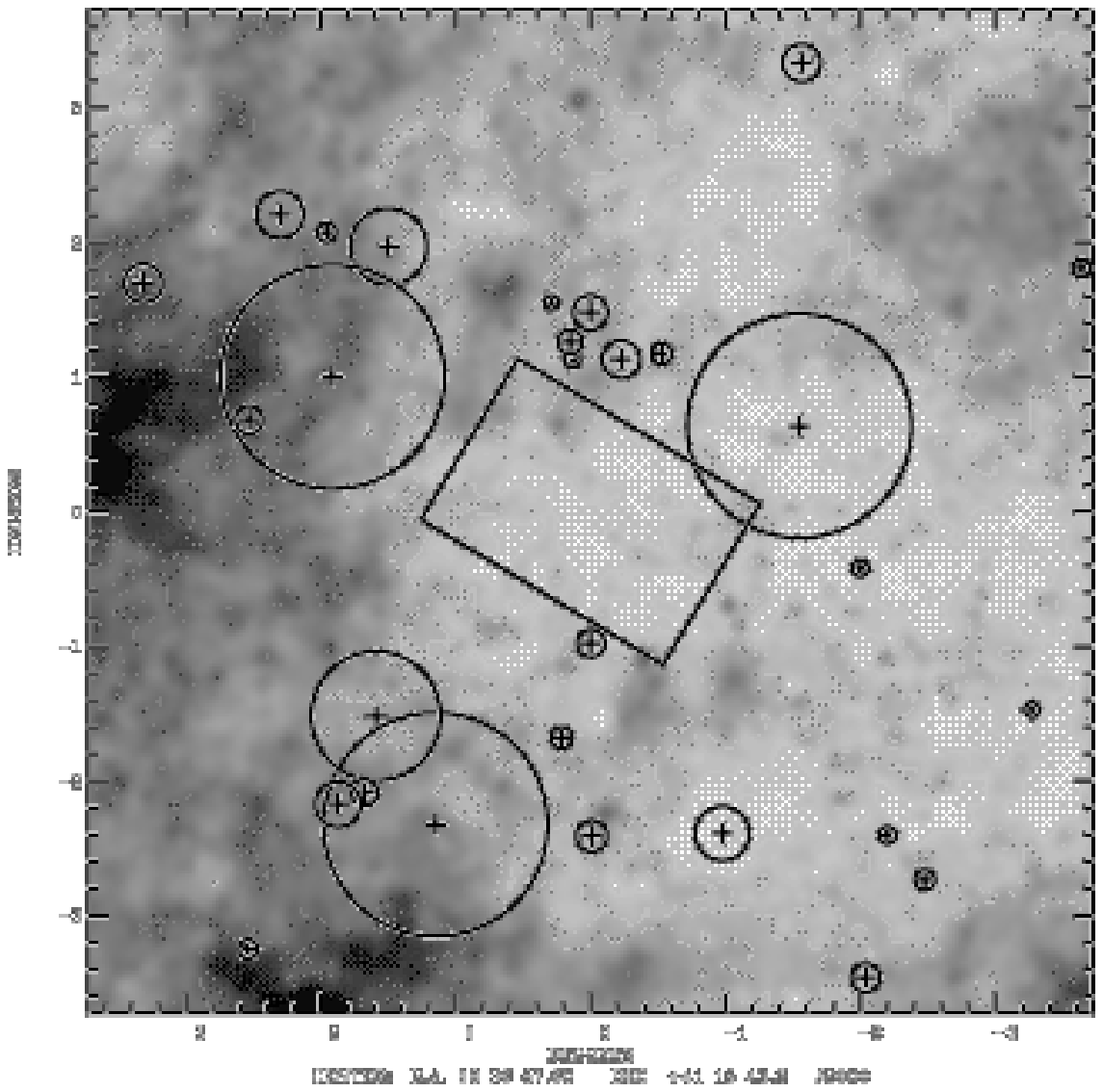,width=8cm}
\caption{The location of the N2 {\em ELAIS} survey
field overlaid on the {\em COBE} normalised {\em IRAS} maps of 
Schlegel et al. (1998).  {\em IRAS} sources with 12\micron\ fluxes brighter than 0.6Jy are
overlayed with radius proportional to flux for $S_{60}<10$Jy. 
 The minimum 100\micron\ intensity shown is 0MJy/sr (white)
and the maximum is 1.5MJy/sr (Black).}\label{fig:i100n2}
\end{figure}

\begin{figure}
\epsfig{file=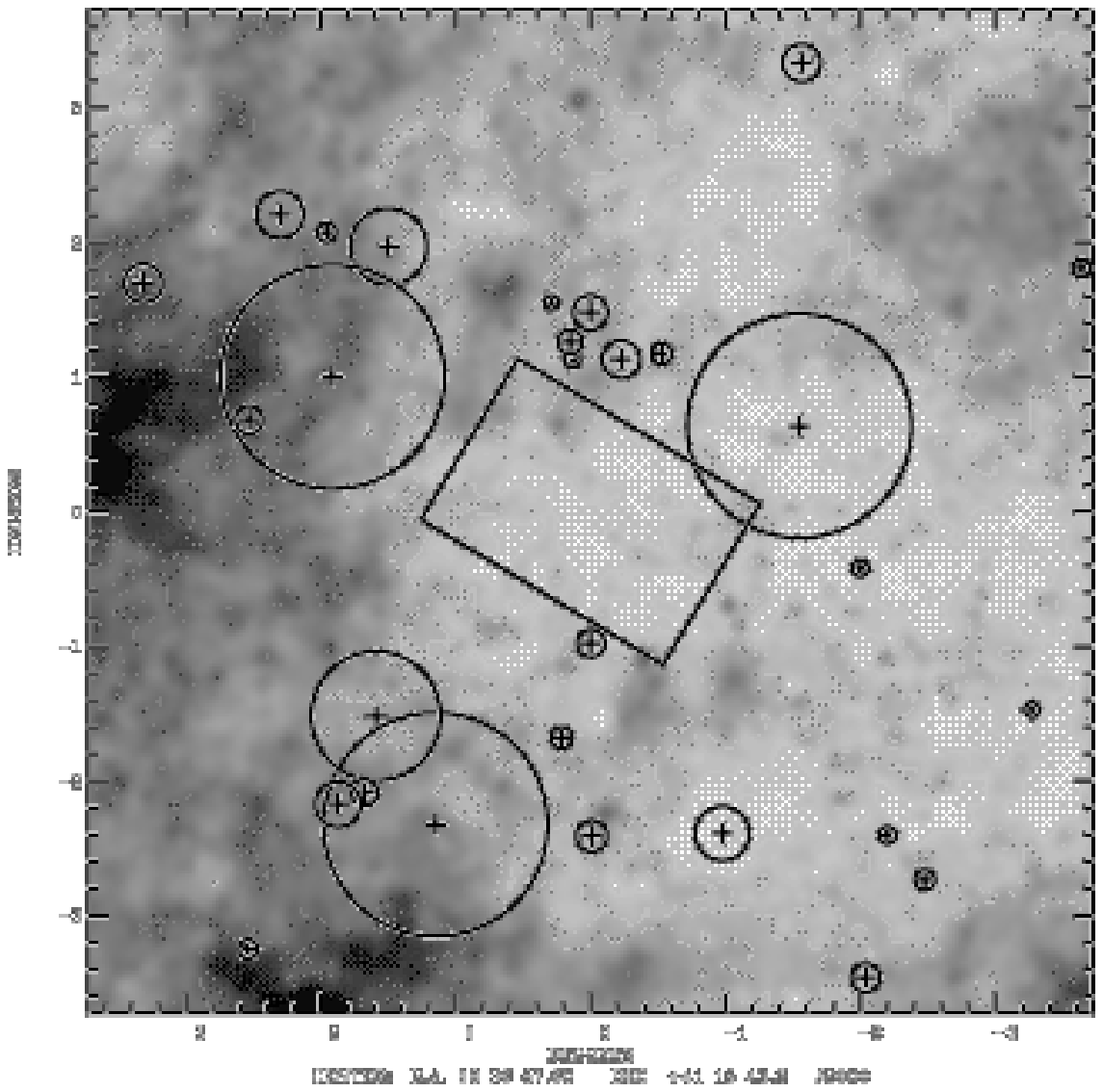,width=8cm}
\caption{The location of the N3 {\em ELAIS} survey
field overlaid on the {\em COBE} normalised {\em IRAS} maps of 
Schlegel et al. (1998).  {\em IRAS} sources with 12\micron\ fluxes brighter than 0.6Jy are
overlayed with radius proportional to flux for $S_{60}<10$Jy. 
The minimum 100\micron\ intensity shown is 0MJy/sr (white)
and the maximum is 1.5MJy/sr (Black).}\label{fig:i100n3}
\end{figure}

\begin{figure}
\epsfig{file=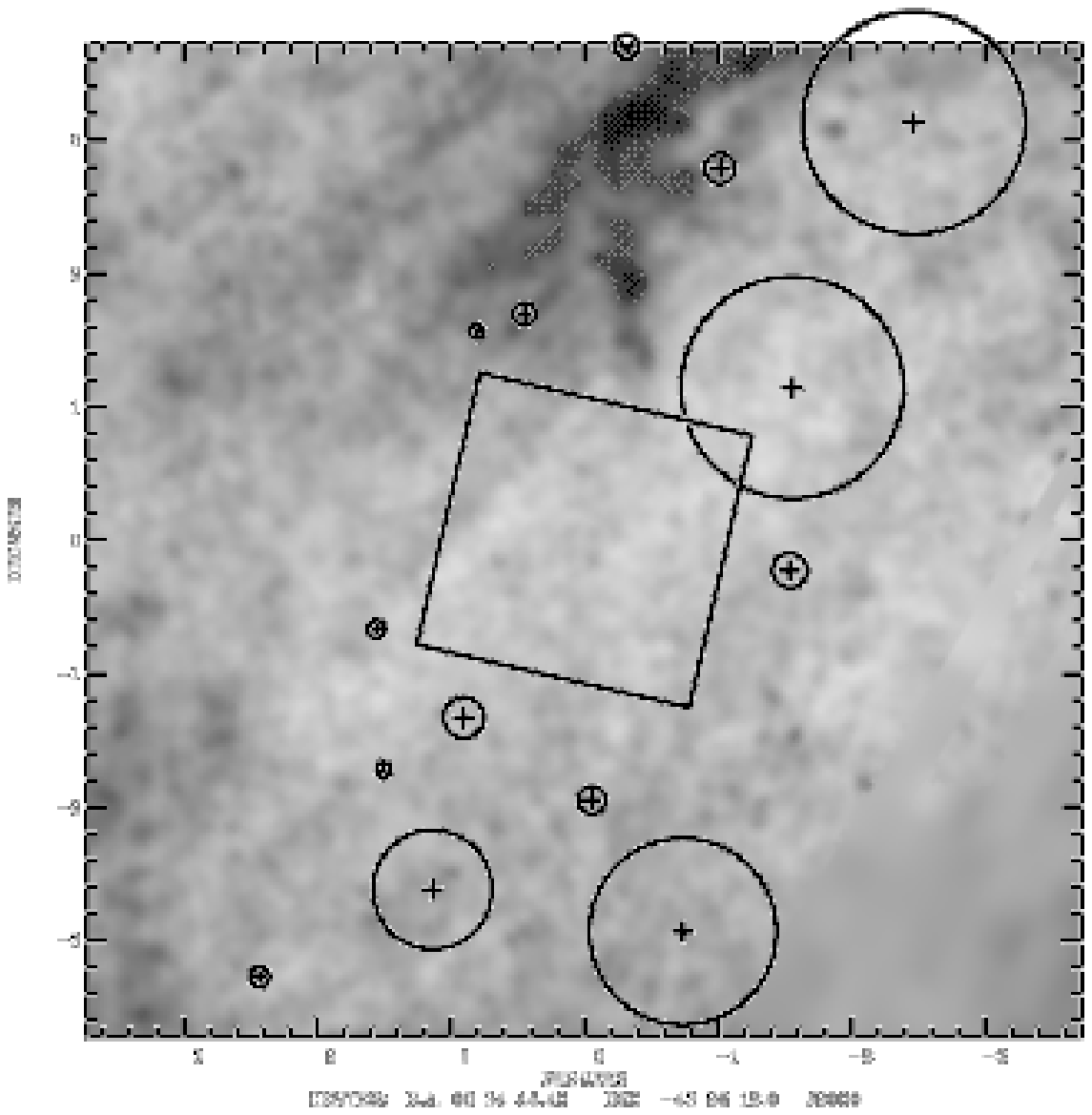,width=8cm}
\caption{The location of the S1 {\em ELAIS} survey
field overlaid on the {\em COBE} normalised {\em IRAS} maps of 
Schlegel et al. (1998).  {\em IRAS} sources with 12\micron\ fluxes brighter than 0.6Jy are
overlayed with radius proportional to flux for $S_{60}<10$Jy. 
The minimum 100\micron\ intensity shown is 0MJy/sr (white)
and the maximum is 1.5MJy/sr (Black).}\label{fig:i100s1}
\end{figure}

The allocated observing time was sufficient to observed around 12
square degrees.  The choice of where to distribute the {\em ELAIS}
rasters on the sky was governed by a number of factors.  Firstly, we
decided not to group these all in a single contiguous region of the
sky; this further reduces the impact of cosmic variance on the survey
(see Section~\ref{survey_lam}).  Distributing the survey areas across
the sky also has advantages for scheduling follow-up work.  Cirrus
confusion is a particular problem, so we selected regions with low
{\em IRAS} 100$\mu$m intensities ($I_{100}<1.5$MJy/sr), using the maps
of Rowan-Robinson et al. \shortcite{Rowan-Robinson et al. 1991b}.  In recognition of the large amount of
time required we decided 
to minimise scheduling conflict with other {\em
ISO} observations by further restricting ourselves to regions of high
visibility ($>25$ per cent) over the mission lifetime, while to reduce
the impact of the Zodiacal background we only selected regions with
high Ecliptic latitudes ($|\beta|>40^\circ$).  Finally, it was
essential to avoid saturation of the {\em ISO-CAM} detectors, so we
had to avoid any bright {\em IRAS} 12$\mu$m sources ($S_{12}>0.6$Jy).
These requirements led us to selecting the four main fields detailed
in the upper portion of Table \ref{tab:fields}. The location of all
{\em ELAIS} fields are indicated in Figure \ref{fig:i100allsky}
showing the Galactic Cirrus distribution, while in Figures
\ref{fig:i100n1}-\ref{fig:i100s1} we show the nominal boundaries of
each of the main survey fields overlayed on a Cirrus map.

Towards the end of the mission an additional field (S2) was
selected with similar criteria,  this field was multiply observed to
provide reliability and completeness estimates. 

A further 6
fields were selected as being of particular interest to warrant a
single small ($24\arcmin\times 24\arcmin$) raster.  These were chosen
either because of existing survey data or because the field contained
a high redshift object and were thus more likely to contain high
redshift {\em ISO} sources. 
\begin{enumerate}
\item{\bf Phoenix:} This  field was the target of a deep radio survey
\cite{Hopkins et al. 1998} and has been extensively followed up from the
ground with imaging and spectroscopy.
\item{\bf Lockman 3:} This  was one of the deep {\em ROSAT}
survey fields \cite{McHardy et al. 1998}.
\item{\bf Sculptor:} This field has been the subject of an extensive
ground based optical survey programme (e.g. Galaz \& De Lapparent
\shortcite{Galaz et al. 1998}).
\item{\bf VLA 8:} This field is centred on a $z=2.394$ radio galaxy
\cite{Windhorst et al. 1991} and has been the target a deep {\em
Hubble Space Telescope} observations \cite{Windhorst et al. 1998}.
\item{\bf TX0211-122:} This object is at $z=2.34$ \cite{van Ojik et al. 1994} and was discovered in the Texas radio
survey \cite{Douglas et al. 1996}.  
\item{\bf TX1436+157:} This object is at $z=2.538$ \cite{Roettgering et al. 1997} and was discovered in the Texas radio
survey \cite{Douglas et al. 1996}. The {\em ISO} field centre is offset from the radio object as the B1950 equinox coordinates were 
entered rather than the J2000 coordinates.
\end{enumerate}
These 6 regions are also described in the
lower portion of Table \ref{tab:fields}.

\begin{table*}

\begin{tabular}{lrrrrrccc}
Name &
\multicolumn{2}{c}{Nominal Coordinates} &
X & Y & ROLL &
 $\langle I_{100}\rangle$ &
 Visibility &
 $\beta$ \\

  & \multicolumn{2}{c}{J2000}
 & /\deg & /\deg & /\deg
 & \begin{small}$/{\rm MJy sr}^{-1}$ \end{small}& /\% & \\
\\
N1&$16^h10^m01^s$&$+54\deg30\arcmin36\arcsec$&
 2.0 & 1.3 & 76 & 0.43 & 98 & 73 \\
N2 &$16^h36^m58^s$&$+41\deg15\arcmin43\arcsec$&
 2.0 & 1.3 & 59 &  0.40 & 59 & 62 \\
N3 &$14^h29^m06^s$&$+33\deg06\arcmin00\arcsec$&
 2.0 & 1.3 & 110 &  0.48 & 27 & 45 \\
S1 &$00^h34^m44^s$&$-43\deg28\arcmin12\arcsec$&
 2.0 & 2.0 & 77 &  0.37 & 32 &-43 \\
\\

S2 &$05^h02^m24^s$&$-30\deg35\arcmin55\arcsec$&
0.3 & 0.3 & 290 & 0.55 & 32 &-43 \\

X1 (Phoenix)& $ 01^h 13^m 13^s $&$ -45\deg 14\arcmin 07\arcsec $ &
0.4 & 0.4 & 33 &  0.62 &  36  & -48 \\
X2 (Lockman 3) &$13^h34^m36^s$&$+37\deg54\arcmin36\arcsec $ &
 0.4 & 0.4 & 280 & 0.28 & 17 & 44 \\
X3 (Sculptor) & $00^h22^m48^s$&$-30\deg06\arcmin30\arcsec $&
0.4 & 0.4 & 254 &  0.99 & 28 & -30 \\
X4 (VLA 8) & $ 17^h 14^m 14^s $&$+50\deg 15\arcmin 24\arcsec $&
0.3 & 0.3 & 162 &  0.87 &  99  &   73 \\
X5 (TX0211-122) & $ 02^h 14^m 17^s $&$-11\deg 58\arcmin 46\arcsec$
& 0.3 & 0.3& 254  & 1.22 & &  -65\\
X6 (TX1436+157) & $14^h36^m43^s$&$+15\deg44\arcmin13\arcsec $&
0.3 & 0.3 & 124 &  1.19 & 22 & 29 \\

 \end{tabular}
\caption{Summary of {\em ELAIS} Survey Fields.
These fields were selected primarily for having low
Cirrus contamination, specifically $I_{100}<1.5$MJy/sr from
the {\em IRAS} maps of Rowan-Robinson et al. \shortcite{Rowan-Robinson et al. 1991b}, the  $I_{100}$ quoted in
this Table are from the maps of Schlegel et al. \shortcite{Schlegel et al. 1998}.
For N1-3, S1 and all X fields we also restricted ourselves to regions of
high visibility $>25$\% over the mission lifetime.  For low Zodiacal background we
required  $|\beta|>40$ and to avoid saturation of
the {\em ISO-CAM} detectors we had to avoid any bright {\em IRAS} 12$\mu$m sources.
Approximate dimensions of the fields in degrees ($X \times Y$) are
given along with the orientation from North to the $Y$ axis Eastwards
in degrees (ROLL).
The 6 smaller rasters X1-6
are centred on well studied areas of the
sky or high-$z$ objects}
\label{tab:fields}

\end{table*}

\subsection{Observation Parameters}\label{obsparam}

Table \ref{tab:surveyaots} summarises the instrument parameters 
specified in the majority of our survey AOTs.  
Most are self explanatory.

For {\em ISO-CAM} the $Gain$ was set to 2 which was the standard
used for most {\em ISO-CAM} observations. $TINT$ was the integration time per 
readout and $NEXP$ was the number of readouts per pointings (i.e. the
total integration time per pointing is $NEXP \times TINT$).  $NSTAB$
was the additional number of readouts for the first pointing
of a raster added to allow the detector to stabilise.

With {\em ISO-PHOT}, $TINT$ was the total integration time per pointing.

The parameters related to the raster geometry
($PFOV$, $NPIX$, $M,N$,  $dM,dN$) have the same meaning
for each instrument. $PFOV$  is the
nominal pixel field of view on the sky.  $NPIX$ is the number of
pixels along each axis of the detector array.  $M,N$ are the number of steps
in a raster while $dM,dN$ are the step sizes.

The {\em ISO-CAM} rasters were designed such that each sky position was
observed twice in successive pointings to improve reliability.   
To reduce overheads we selected  a very large raster size, $40\arcmin\times40\arcmin$. 
With the exception of small rasters and one test raster, the {\em ISO-CAM} 
parameters remained unchanged throughout the survey.

Since the {\em ISO-PHOT} internal calibration measurements were only performed 
at the beginning and end of a raster we chose these to be half the
size of the {\em ISO-CAM} rasters ($20\arcmin\times40\arcmin$).
We originally  used  {\em ISO-PHOT} with a non-overlapping raster pattern  and
switched to an overlapping mode
during the mission, with  most  N1 and S1 observations performed in
the non-overlapping mode.

The observation parameters for all survey observations are tabulated
in Appendix \ref{isolog}.

\begin{table}
\begin{tabular}{lllll}
   & \multicolumn{2}{c}{{\em ISO-CAM}} & \multicolumn{2}{c}{{\em ISO-PHOT}} \\
\\
Parameter      & &&&\\
Detector        & LW    & LW   & C100  & C200 \\
Filter          & LW2   & LW3  & C90   & C160 \\
$\lambda/ \mu$m   & 6.75   &  15  &  95.1   & 174  \\
$\Delta\lambda/ \mu$m  &   3.5    &  6   &  51.4     &  89.4    \\
$Gain$            & 2     &   2  &  n/a  & n/a  \\
$TINT$/s          & 2     & 2    &  20   & 32   \\
                  &       &      &  12   &      \\
$NEXP$            & 10    & 10   & n/a   & n/a  \\
$NSTAB$           & 80    & 80   & n/a   & n/a  \\
$PFOV$/$^{\prime\prime}$    & 6     & 6     & 43.5 & 84.5 \\
$NPIX$            & 32    & 32   & 3     & 2 \\
$M,N$           & 28, 14& 28, 14& 10, 20 & 13, 13 \\
                &       &      &  20, 20 &      \\
$dM,dN$/$^{\prime^\prime}$ & 90, 180 & 90,180  &  130, 130 & 96, 96 \\
                &       &      &  75, 130 &      \\
\end{tabular}
\caption{Summary of the AOT parameters for the bulk of
survey programme observations.  The 90\micron\ survey
strategy changed after the S1 and N1 observations, and
the revised parameters are illustrated.  Many observations
in N3 and all of the smaller fields  X1-6 and S2 were executed
with very similar AOTs, but with fewer pointings. Parameters
are described in the text and in detail on the {\em ISO} WWW pages
(http://isowww.estec.esa.nl/). 
}\label{tab:surveyaots}
\end{table}

\section{{\em ISO} Observations}

The {\em ISO} observations for the {\em ELAIS} programme were executed from
12th March 1996 (revolution 116), 37 days after the
beginning of routine operations (4th February 1996, revolution 79)
 until 17:44 on 8th April 1998
(revolution 875), 10 hours 44 minutes after the first signs of
boil off had been detected and 5 hours 23 minutes before the
last observations were performed.

In general the execution of the planned observations was
very successful.  Only three observations were reported as
``failed''.   Three observations were flagged as ``aborted'',
all three of these had been  concatenated to ``failed'' observations
but appear to have been successfully executed despite this.

The only significant problem in the execution of the survey
observations occurred in  N3.  It transpired that there was
a paucity of guide stars in this region and the mission planning
team were unable to schedule many of the observations near
the original dates requested.  To accommodate this
problem the sizes of the rasters
were reduced and restrictions on the possible observation dates relaxed.
However, in the last available observing window for N3
 other {\em ISO} mission priorities, 
together with
remaining guide star acquisition problems, interfered
with the scheduling.  The net result is that the coverage
of the N3  region is patchy.  

It may be that this guide star problems noticed in N3 may be related
to an apparent offset of around 6'' between the reference frame of the 
DSS and e.g. the APM catalogue in this field.  The APM catalogue
agrees very well with the Guide Star Catalogue v1.2 
(\verb+http://www-gsss.stsci.edu/gsc/gsc12/gsc12_form.html+).

\subsection{Main Survey Observations}

Table \ref{tab:field_areas} indicates the area that has been
surveyed at least once in any band in all of our fields.  For the four large
fields the separation of the raster pointings (40\arcmin)
is used to compute the area, i.e. 0.44 square degrees
per raster.  For the small fields which are not mosaiced
the actual size of the raster is used.

\begin{figure*}
\epsfig{file=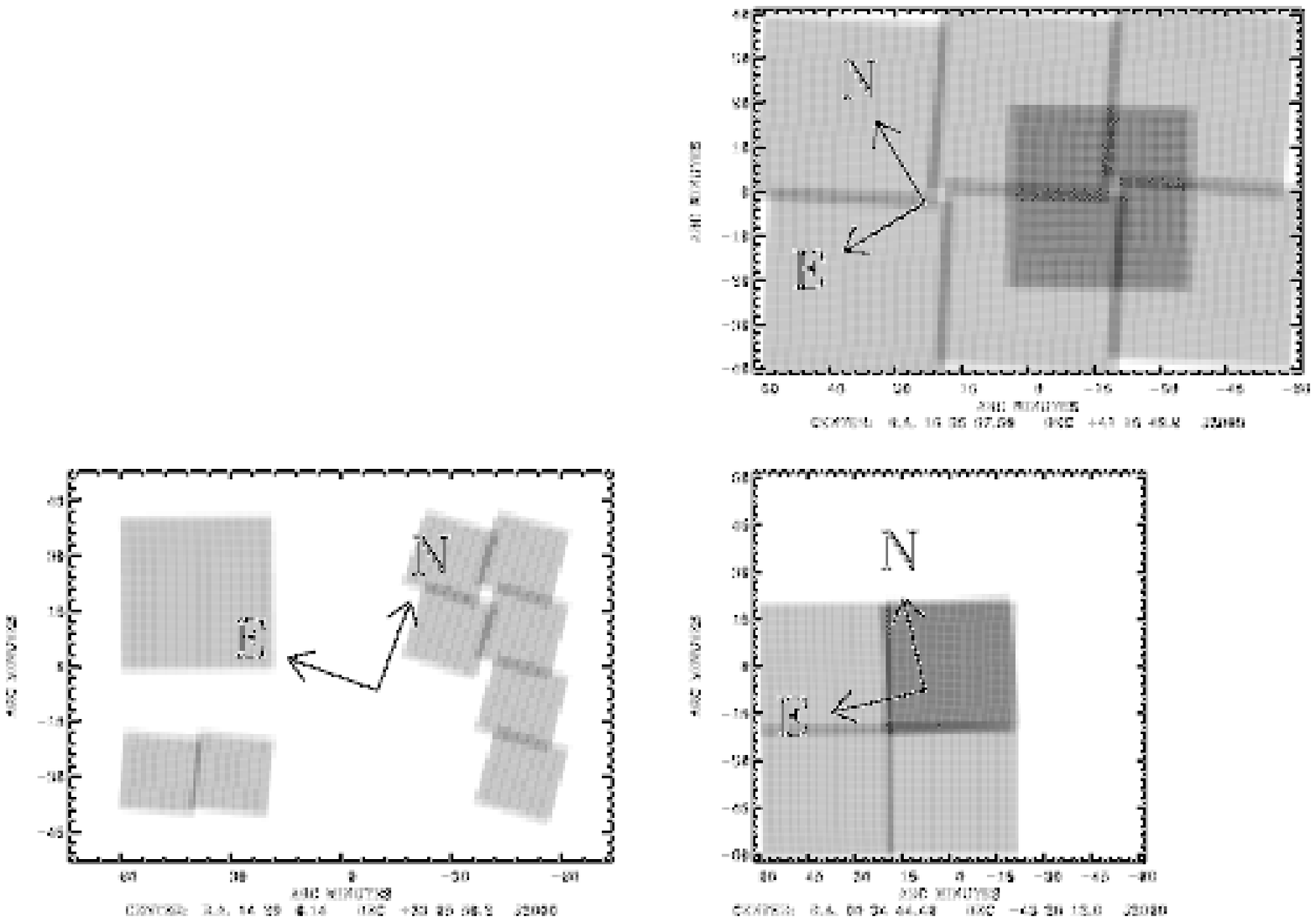}
\caption{Survey Coverage at 6.7\micron\. White 
areas  have not been covered at all darker regions indicate longer
total integration, either due to repeated observations or overlap,
black indicates 200s. The coverage maps have been smoothed to one arc
minute resolution. Reading from top left to
bottom right the fields are: N2, N3, S1. The true peaks in the
coverage are around 150s, 100s and 175s respectively.
}\label{fig:cov7}
\end{figure*}

\begin{figure*}
\epsfig{file=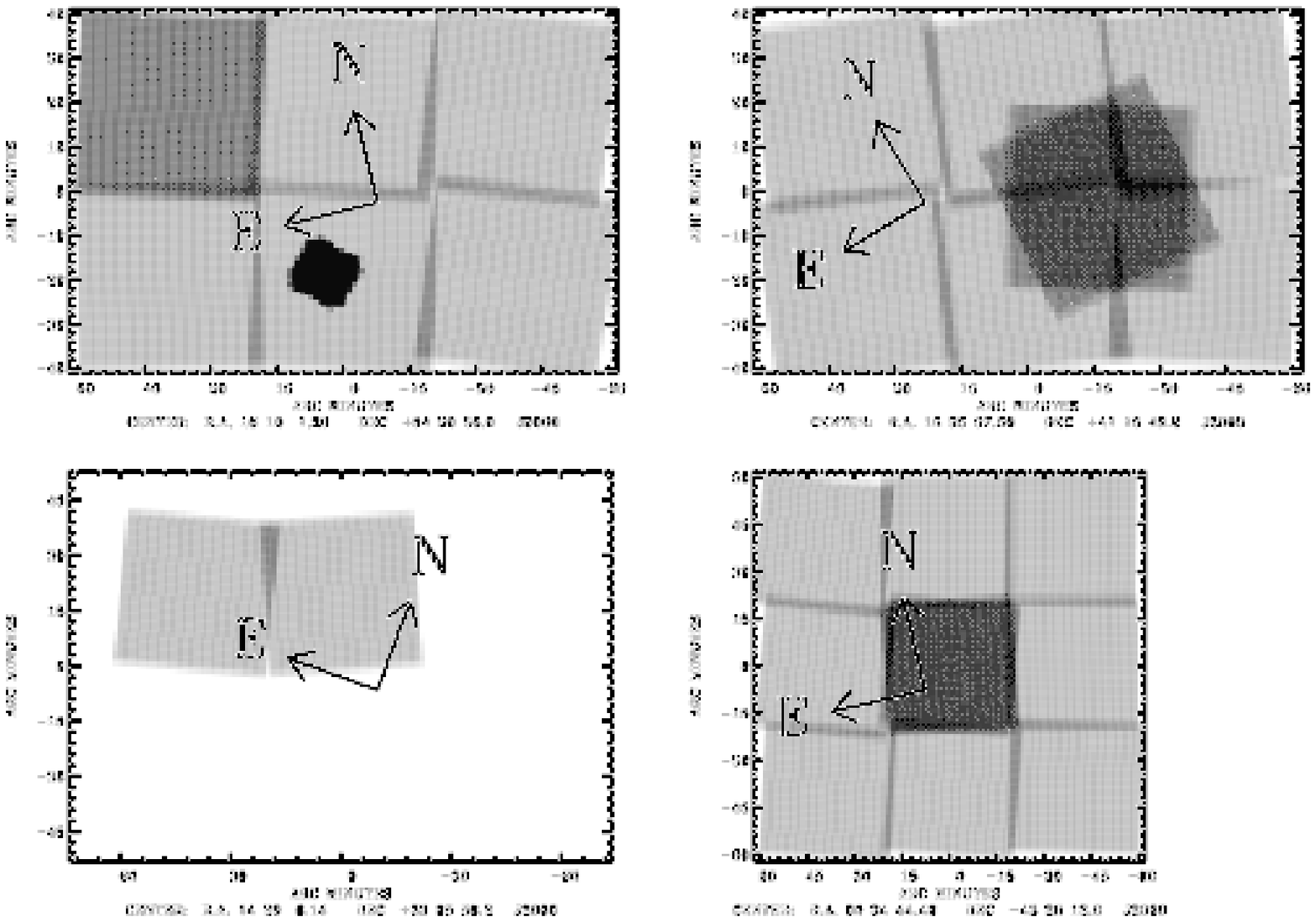}
\caption{Survey Coverage at 15\micron\. White 
areas  have not been covered at all darker regions indicate longer
total integration, either due to repeated observations or overlap,
black indicates 200s. The coverage maps have been smoothed to one arc
minute resolution. Reading from top left to
bottom right the fields are: N1, N2, N3, S1. The true peaks in the
coverage are around 600s, 200s, 100s and 200s respectively.
}\label{fig:cov15}
\end{figure*}

\begin{figure*}
\epsfig{file=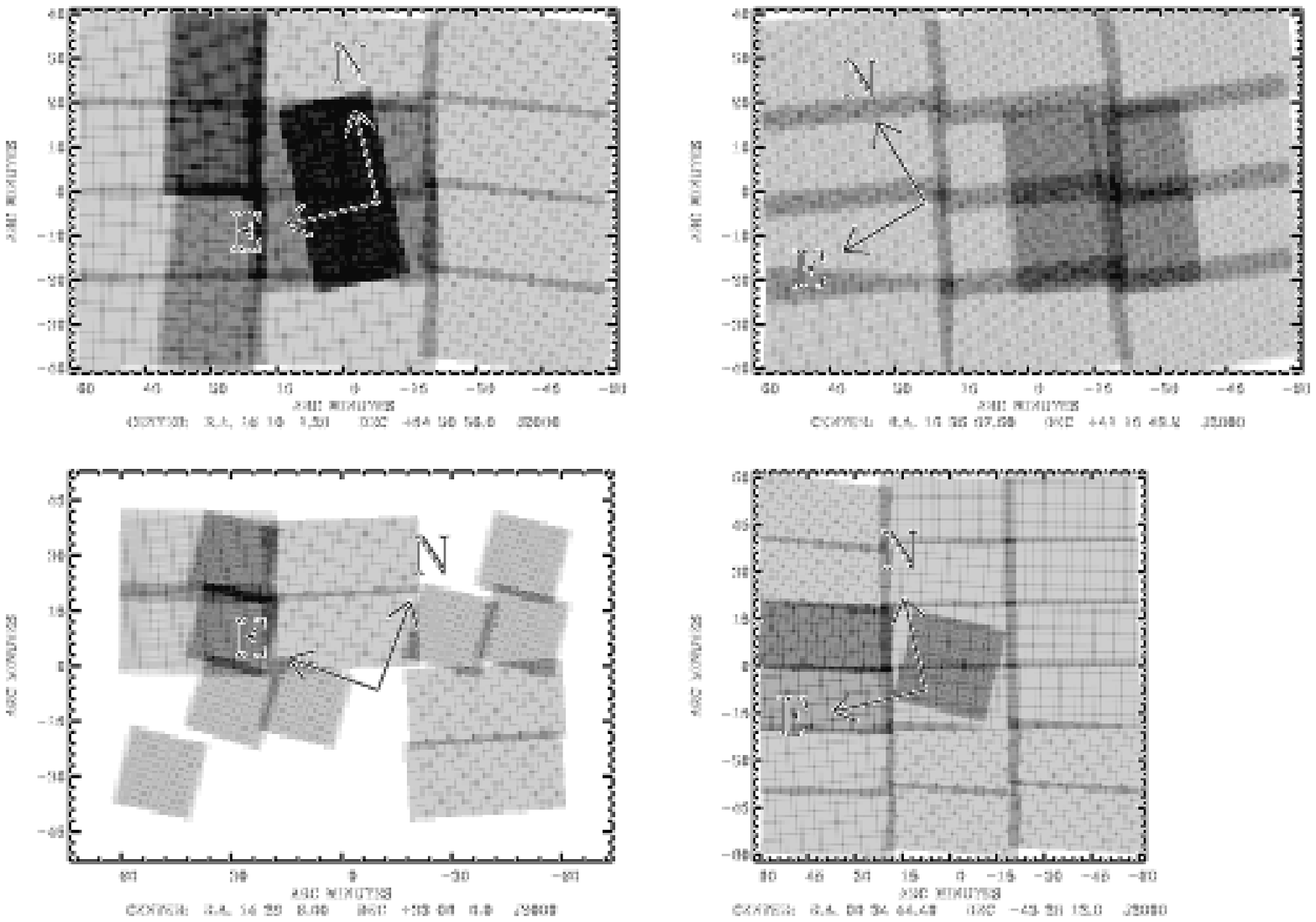}
\caption{Survey Coverage at 90\micron\. White 
areas  have not been covered at all darker regions indicate longer
total integration, either due to repeated observations or overlap,
black indicates 100s. The coverage maps have been smoothed to one arc
minute resolution. Reading from top left to
bottom right the fields are: N1, N2, N3, S1. The true peaks in the
coverage are around 160s, 120s, 120s and 120s respectively.
}\label{fig:cov90}
\end{figure*}

\begin{figure*}
\epsfig{file=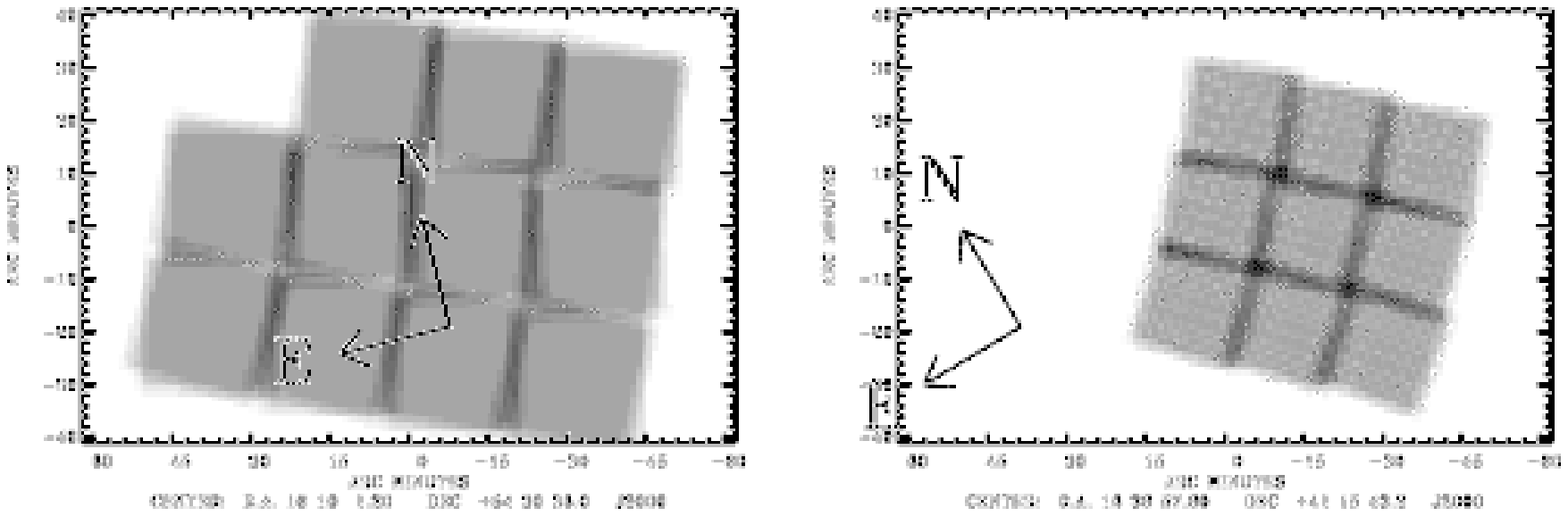}
\caption{Survey Coverage at 175\micron\.  N.B. Only the coverage in
N2 is part of the {\em ELAIS} project, the coverage in N1 (left) comes from the
FIRBACK team (PI J-L. Puget, see Dole et al. \shortcite{Dole et al. 1999}). White 
areas  have not been covered at all darker regions indicate longer
total integration, either due to repeated observations or overlap,
black indicates 350s. The coverage maps have been smoothed to one arc
minute resolution.  The true peaks in the
coverage are around 260s and 350s respectively.
}\label{fig:cov175}
\end{figure*}

The coverage, in terms of integration time per sky pixel, of the four main survey fields (N1-3 and S1)
in each of the bands are shown in Figures \ref{fig:cov7},
\ref{fig:cov15},\ref{fig:cov90},\ref{fig:cov175}.

\begin{table}
\begin{tabular}{ccccc}
Field & \multicolumn{4}{c}{Wavelength/\micron} \\
     & 6.7     & 15     & 90    & 175 \\
     &       &        &       &     \\
N1   &       &  2.67  &  2.56 & 2$^\star$    \\
N2   & 2.67  &  2.67  &  2.67 & 1    \\
N3   & 1.32  &  0.88  &  1.76 &     \\
S1   & 1.76  &  3.96  &  3.96 &     \\
S2   & 0.12  &  0.12  &  0.11 &  0.11   \\
X1   &       &  0.16  &  0.19 &     \\
X2   &       &  0.16  &  0.19 &     \\
X3   &       &  0.16  &  0.19 &     \\\hline
     & 5.87  & 10.78  & 11.63 &  3.11 \\ \hline
X4   &       &  0.09  &  0.11  &     \\
X5   &       &  0.09  &        &     \\
X6   &       &  0.09  &  0.11  &     \\
\end{tabular}
\caption{Survey Fields covered at least once.   Areas are given in
square degrees.$^\star$The 175\micron\ observations in N1 have been 
carried out by the FIRBACK team (PI J-L Puget, see Dole et
al. \shortcite{Dole et al. 1999})
and are included in this table to illustrate the complete {\em ISO} coverage
of the {\em ELAIS} fields.  N1-3, S1-2 and X1-3 are unbiased survey
fields, while X4-X6 are centred on known objects so should not be
included with the other fields for statistical purposes.
}\label{tab:field_areas}
\end{table}

\subsection{Duplicate Survey Observations}

A number of sub-fields have been repeated on one or more occasions.  This
repetition will considerably aid in assessing the reliability and
completeness of the survey.  In addition this data will provide deeper
survey regions which are good targets for more focussed follow-up
campaigns and other exploitation.

Table \ref{tab:duplicates} lists all the fields that have repeated
observations together with the level of redundancy.

\begin{table}
\begin{tabular}{lllcccc}
Field & \multicolumn{2}{c}{Coordinates (J200)}&\multicolumn{4}{c}{Wavelength/\micron} \\
     & &&6.7 & 15 & 90 & 175 \\
     & &&  &    &     &     \\
N1\_T &$16^h10^m01^s$&$+54\deg30\arcmin36\arcsec$  
 & &   &  0.22$\times$3 & 2$\times$2\\
      & & 
 &   &     &   0.22$\times$4  & \\
N1\_U &$16^h11^m00^s$&$+54\deg13\arcmin23\arcsec$  
 & & 0.04$\times$11    &  & \\
   
N1\_1C& $16^h13^m54^s$&$+54\deg18\arcmin22\arcsec$  
 & & & 0.22$\times$2 &  \\
N\_2  &$16^h13^m57^s$&$+54\deg59\arcmin36\arcsec$  
 &  0.44$\times$2 & &     \\

N2   &$16^h35^m45^s$&$+41\deg06\arcmin00\arcsec$ 
& 0.44$\times$2  & 0.44$\times$3   &   0.44$\times$3     &   1$\times$2    \\
N3\_5C/D   &$14^h31^m53^s$&$+33\deg14\arcmin37\arcsec$  &    &   0.22$\times$2  &     \\
S1\_5  &$00^h34^m44^s$&$-43\deg28\arcmin12\arcsec$& 
 0.44$\times$2   &  0.44$\times$3   &  0.22$\times$3    &     \\
S2   &$05^h02^m24^s$&$-30\deg35\arcmin55\arcsec$
& &    0.12 $\times$4 &  0.11$\times$5   & 0.11$\times$3 \\
\end{tabular}
\caption{Survey sub-fields covered more than once, listing the area
covered (in square degrees) and the 
number of times that sub-field has been observed.}\label{tab:duplicates}
\end{table}

\subsection{Photometry Programme}

In addition to the survey observations, we also undertook a
photometry programme to observe objects detected early on
in the survey programme at other {\em ISO} wavelengths.

These objects were selected from the S1 and N1 survey regions
which had been observed at an early stage in the campaign.
180 objects which had been detected at 15 $\mu$m were selected
to be observed with {\em ISO-CAM} at 4.5, 6.7, 9, and 11 $\mu$m using
the filters LW-1,LW-2,LW-4 and LW-7.   80 Objects were selected to be
observed by {\em ISO-PHOT} at 60 and 175 $\mu$m, using the C60, and C160
filters.

The {\em ISO-CAM} observations were performed in concatenated
chains of 10 pointings.  At each pointing a $2\times 1$ raster was
performed to ensure accurate photometry and reliable detections.
The chains were arranged such that each of the 10 sources was located in
a different position on the array (separated by around 18\arcsec),
this was to allow accurate sky flat-fielding over the course of the
concatenated chain.  The 120 {\em ISO-CAM} pointings in S1, and the 80
pointings in N1 were ordered to minimize the total path length,
ensuring that sequential observations were as close to
each other as possible, both spatially and temporally, improving
the flat-fielding.

The {\em ISO-PHOT} observations were performed in chains of 15 pointings.
On average the 15 pointings contained 5 source positions and 10
background positions.  Like the {\em ISO-CAM} photometry observations, the
{\em ISO-PHOT} source positions were ordered to minimize the total path
length.  The background pointings were chosen to be
spaced along this path at reasonably regular intervals, 
while ensuring that there was at least one background position between every source position.

Other parameters from the AOTs for the photometry programme
are summarised in Table \ref{tab:photomaots}.

\begin{table}
\begin{tabular}{lllllll}
Instrument   & \multicolumn{4}{c}{{\em ISO-CAM}} & \multicolumn{2}{c}{{\em ISO-PHOT}} \\
\\
Parameter       & &&&&&\\
\\
Detector        & LW  & LW  & LW  & LW  & C100 & C200 \\
Filter          & LW1 & LW2 & LW4 & LW7 & C60 & C90   \\
$\lambda/ \mu$m   & 4.5 & 6.7 &   6.0 & 9.62  & 60.8  & 174   \\
$\Delta\lambda/ \mu$m  &  1   &  3.5   & 1    &  2.2   &    23.9 &  71.7     \\
$Gain$            & 2   & 2   &   2 &   2 &  n/a & n/a  \\
$TINT$/s          & 5   & 5   & 5   & 5   &  32  & 32   \\
$NEXP$            & 8   & 8  & 8  & 8  & n/a  & n/a  \\
$NSTAB$           & 50  & 50  & 50  & 50  & n/a  & n/a  \\
$PFOV/^{\prime\prime}$    & 6   & 6   & 6   & 6   & 43.5 & 84.5 \\
$M,N$           & 2,1 & 2,1 & 2,1 & 2,1 & 1,1  & 1,1  \\
$dM,dN/^{\prime\prime}$ & 24  & 24  & 24  & 24  & n/a  & n/1  \\

\end{tabular}
\caption{Summary of the AOT parameters for the photometry
programme}\label{tab:photomaots}
\end{table}

\section{Data Processing and Products}

In order to provide targets early on in the campaign to allow
follow-up programmes, both from the ground and with {\em ISO}, it was
decided to perform an initial ``Preliminary Analysis''. This was
started long before the end of the mission, while the understanding
of the behaviour of the instruments was naturally less than it is
currently and will be superseded  with a ``Final Analysis''
incorporating the best available knowledge post mission.

The Preliminary Analysis  was conducted
with the intention  of producing reliable source lists at $6.7, 15$ and 
$90 \mu$m.   The processing of the {\em ISO-CAM} survey observations 
is  described in detail by Serjeant et al. \shortcite{Serjeant et al. 1999}
and the reduction of the {\em ISO-PHOT} 90$\mu$m survey data will
be discussed by Efstathiou et al. (1999, in preparation).

The Final Analysis is currently being undertaken.   This
is expected  to produce better calibrated and  fainter source lists than  the Preliminary
Analysis.  The Final Analysis will also produce maps which can be used
to determine fluxes or upper-limits for known sources. This analysis
will not, however,  be completed until early 2000.

The {\em ELAIS} products will  comprise source catalogues at all
wavelengths, 4.5, 6.7, 9, 11, 15, 60, 90, 175, together with
maps from all the survey observations.  Highly reliable sub-sets of
the ``Preliminary Analysis'' catalogues were released to the community,
via our WWW site (\verb+http://athena.ph.ic.ac.uk/+), concurrent with the expiration
of the propriety period on 10th August 1999.

\subsection{Data Quality}

The quality of the 15 \micron\ {\em ISO-CAM} data is moderately uniform.  Some
rasters are more affected by cosmic rays than others but the total
amount of data seriously affected by cosmic rays is small. 
The noise levels are within a factor of a few
of those expected;  a typical noise level is 0.2 ADU/sec/pixel per pointing.

The {\em ISO-PHOT} data is seriously affected by cosmic rays and detector
drifts.  We have used the fluctuations in the time sequence of each
pixel as an estimate of the average noise level. 
The fluctuations per pointing were typically 3 per cent of the background level,
though 3 of the 9 pixels were noisier with fluctuations typically 4 per cent of the background.
 A few observations showed
higher noise due to increased cosmic ray hits.  Our original AOTs
employed an integration time of 20s.  We subsequently decreased this
to 12s to allow for an overlapping raster giving a factor of two
redundancy with similar observation time.  Importantly there does not appear
to be a significant difference in noise levels per pointing despite
the factor of two reduction in integration time, indicating that
non-white noise in the pixel histories is dominant.  
The redundancy introduced by this new strategy could 
improve the signal to noise ratio for sources by as much as $\sqrt{2}$.

\subsection{Preliminary Data Analysis}

The processing for the {\em ISO-PHOT} and {\em ISO-CAM} data proceeds in a similar
fashion.  All data reduction used a combination of standard routines
from the {\em PHOT} Interactive Analysis \cite{pia}
\footnote{PIA is a joint development by the ESA Astrophysics Division
and the {\protect\em ISO-PHOT} consortium})
software and the {\em CAM} Interactive Analysis \cite{cia}
together with purpose-built IDL routines.  The frequency of
glitches and other transient phenomena led to non-Gaussian and
non-white-noise behaviour.  

A number of data reduction techniques were tested at ICSTM,
CEA/SACLAY, IAS and MPIA.  Parallel
pipeline processes for reducing the  {\em ISO-PHOT} data were run at
both ICSTM and MPIA.  Data reduction techniques suitable for {\em
ISO-CAM} data with multiple redundancy, such as the observations of
the Hubble Deep Field \cite{Serjeant et al. 1997},  e.g. the Pattern
REcognition Technique for {\em ISO-CAM} data \cite{Aussel et al. 1999} were unsuccessful in processing this data.
The most reliable approach for
source extraction was found to be looking for source profiles
in the time histories of individual pixels rather than by constructing
sky maps.

For both instruments the
data stream from each detector pixel was treated as an independent scan
of the sky.  These data streams were filtered to remove glitches and
transients and averaged to produce a single measurement at each
pointing position.  Significant outliers remaining in the data
streams were flagged as potential sources.  For the {\em ISO-CAM} observations
the redundancy of the pointings was used to provide confirmation of
candidate sources.  The
data stream surrounding all remaining candidates was then examined
independently by at least two observers to remove spurious detections.
Sources that were acceptable to two or more observers were classified
as good ($REL=2$) and those acceptable to only one observer were
classified as marginal $(REL=3)$.

The fraction of spurious detections
was high due to the non-Gaussian nature of the noise and relatively
low thresholds applied. More than 13 thousand {\em ISO-PHOT} source candidates
were examined as were just over 15 thousand {\em ISO-CAM} 15 \micron\ candidates.
At 6.7  \micron\ the rejected fraction was lower and the candidate
list was only 3
thousand.  The final numbers of objects in
the Preliminary Catalogue Version 1.3 are tabulated in 
Table \ref{tab:qla_cat}

\begin{table}
\begin{tabular}{lrrr}
                   & \multicolumn{3}{c}{Wavelength}\\
Quality            &   6.7$\mu$m & 15 $\mu$ & 90$\mu$\\
Good   $(REL=2)$   &  795     &  728     & 153    \\
Moderate $(REL=3)$ &  2341     &  818     & 208    \\
\end{tabular}
\caption{{\em ELAIS} Preliminary Source Catalogue Statistics}\label{tab:qla_cat}
\end{table}

The ``eye-balling'' technique while laborious ensured that the
resulting catalogues are highly reliable, as discussed in greater
detail in  Serjeant et al. \shortcite{Serjeant et al. 1999} and Efstathiou et al. (1999., in
preparation). The sub-sets of the Preliminary
Catalogues that were released to the community were those {\em
ISO-PHOT} sources that had been confirmed by four observers, and those
{\em ISO-CAM} sources that had been confirmed by two observers with
fluxes above 4mJy, these sub-sets are exceptionally reliable.

A ``Final Analysis'' process has been developed which
uses the transient correction techniques of Lari (1999, in
preparation).  These techniques have been shown to be excellent for
reducing {\em ISO-CAM} data.  While  this is almost certainly the best
procedure for reducing the {\em ELAIS} data, it is labour intensive
and time consuming and we do not expect the ``Final Analysis'' to be
finished until early 2000, hence the release of our ``Preliminary''
products.

\subsection{Source Calibration}

For the {\em ISO-CAM} observations we have of order 10 stars per
raster and these provide a very good calibration.  A preliminary analysis
of the star fluxes (Crockett et al. 1999 in preparation, see also Serjeant et al. 1999) 
suggests that our raw instrumental units (ADU/g/s) need
to be multiplied by a factor of 1.75 to give fluxes in mJy.  This implies a
$50$ per cent completeness limit of approximately $3$ mJy at $15\mu$m. The flux
calibration is still uncertain at $6.7\mu$m, due largely to the uncertain
aperture corrections to the under-sampled observations and the single-pixel
detection algorithm, though PSF models and pre-flight sensitivity
estimates suggest a $50$ per cent completeness level at less than $1$mJy. (See
Serjeant et al. \shortcite{Serjeant et al. 1999} for more details.)

For the 90 \micron\ survey the calibration proceeded as follows.
The expected background was  estimated 
using {\em COBE} and {\em IRAS} data and Zodiacal light
models.  These predictions were compared to the measurements of the
background calibrated using the internal calibration device (FCS) 
allowing  the predicted backgrounds to be corrected from an
extended source to a point source calibration.  These predictions
were then used to scale the measured fluctuations above
the background.
Single pixel detections
(``Point sources'') were then calibrated using the expected fraction
of flux falling on a single pixel for a source placed arbitrarily with
respect to the pixel centre.  The fluxes of ``extended sources'' were
calculated in a more complicated fashion and have great associated
uncertainties.  The fluxes were found to be in good agreement
with model stellar fluxes in our own dedicated calibration
measurements and with the fluxes of {\em IRAS} sources in the fields.
This suggests a 5$\sigma$ noise level of  100mJy.  This {\em
ISO-PHOT} calibration, completeness and reliability estimates is 
discussed in  detail by Efstathiou et al. (1999, in preparation) and Surace et al.
(1999, in preparation).

\section{Comparison with Other {\em ISO} Surveys}\label{isosurveys}

\begin{figure}
\epsfig{file=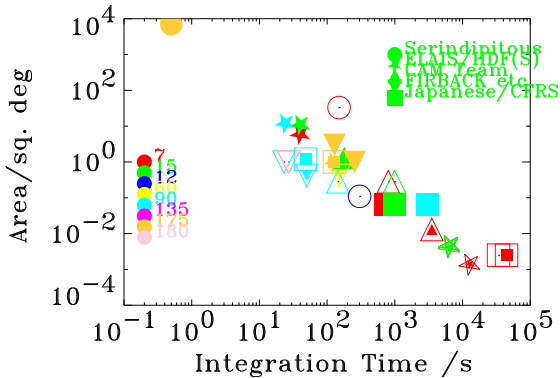,width=8cm}
\caption{Figure comparing {\em ISO} surveys area vs. depth.  All the
surveys in Table \protect{\ref{surveys}} are plotted.  The {\em ELAIS}
points have an additional outline.}
\label{fig:iso_surveys}
\end{figure}

 {\em ISO} carried out a variety of complementary surveys exploring 
the available 
parameter space of depth and area.  Table \ref{surveys} summaries
the main extra-galactic blank-field surveys.  With the exception of
the two main  serendipity surveys {\em ELAIS} covers the
largest area and has produced the largest number of {\em ISO} sources.
Figure \ref{fig:iso_surveys} illustrates how  deeper smaller area
surveys are complemented by shallower wider area surveys.

\begin{table*}
\begin{center}

\begin{tabular}{lllll}
\hline
Survey Name  & [e.g. ref] & Wavelength & Integration & Area\\
             &    &   $/\mu$m  &   $/$s      & $/{\rm sq deg}$\\
\\
PHT Serendipity Survey 	& 1  & 175         & 0.5            & 7000 \\
CAM Parallel Mode 	& 2    & 6.7           & 150            & 33 \\
{\em ELAIS}         	& 3& 6.7,15,90,175 & 40, 40, 24, 128& 6, 11, 12,1\\
CAM Shallow   		& 4      & 15          & 180            & 1.3  \\
FIRBACK      		& 5& 175         & 256, 128       & 1, 3  \\
IR Back       		& 6      & 90, 135,180 & 23, 27, 27     & 1, 1, 1 \\
SA 57         		& 6.7& 60, 90      & 150, 50        & 0.42,0.42  \\
CAM Deep      		& 8 & 6.7, 15, 90   & 800, 990, 144  & 0.28, 0.28, 0.28\\
Comet fields       	& 9& 12          & 302            & 0.11\\
CFRS          		& 10                   & 6.7,15,60,90  & 720, 1000, 3000,3000 & 0.067.0.067.0.067,0.067\\
CAM Ultra-Deep		& 11      & 6.7           & 3520           & 0.013 \\
ISOHDF South  		& 12         & 6.7, 15       &$>6400, >6400$  & 4.7e-3, 4.7e-3 \\
Deep SSA13              & 13               & 6.7           & 34000          & 2.5e-3\\
Deep Lockman            & 14,15      & 6.7, 90, 175  & 44640, 48, 128 & 2.5e-3, 1.2 , 1 \\
ISOHDF North            & 15       & 6.7, 15       & 12800, 6400    & 1.4e-3, 4.2e-3 \\
\hline
\end{tabular}
\end{center}
\caption{Field Surveys with {\protect\em ISO}, ordered roughly in decreasing
area. References: 
1 -- Bogun et al. \shortcite{Bogun et al. 1996}, 
2 -- Siebenmorgen et al.  \shortcite{Siebenmorgen et al. 1996}, 
3 -- this paper, 
4 -- Elbaz et al. \shortcite{Elbaz et al. 1999}, 
5 -- Dole et al. \shortcite{Dole et al. 1999},
6 -- Mattila et al. (1999, in prep.), 
7 -- Linden-V{\o}rnle\shortcite{Linden-vornle 1997}, 
8 -- Elbaz et al. \shortcite{Elbaz et al. 1999}
9 -- Clements et al. \shortcite{Clements et al. 1999},
10 -- Flores et al. \shortcite{Flores et al. 1999a},\shortcite{Flores et al. 1999b}, 
11 -- Elbaz et al. \shortcite{Elbaz et al. 1999}, 
12 -- Oliver et al. (1999, in preparation), 
13 -- Taniguchi et al. \shortcite{Taniguchi et al. 1997a}, 
14 -- Taniguchi et al. \shortcite{Taniguchi et al. 1997b}, 
15 -- Kawara et al. \shortcite{Kawara et al. 1998}, 
16 -- Serjeant et al. \shortcite{Serjeant et al. 1997}}\label{surveys}
\end{table*}

\section{Follow-up}\label{followup}

An extensive follow-up programme is being undertaken,
including observations in many bands from X-ray to radio.  This
programme will provide essential information for identifying the
types of objects detected in the infrared, their luminosities,
energy budgets and other detailed properties.  As well as
studying the properties of the objects detected by {\em ISO} a number
of the follow-up surveys will provide independent source lists
which will be extremely valuable in their own right,  e.g. to
investigate the differences between infrared and non infrared emitting objects.

\subsection{Surveys}
A number of follow-up programmes are in fact independent surveys at
other wavelengths, carried out within the {\em ELAIS}
survey area. These include:

\begin{enumerate}

\item {\bf Optical:}
$R$-band CCD surveys are essential to provide optical identifications for
spectroscopic and related follow-up with  improved astrometry,
photometric accuracy and to fainter levels than those provided by the Second Sky
Survey. 
Our principal southern field (S1) has been completely covered with the ESO/Danish
1.5m telescope to a depth
of  $R\sim 23.5$ (La Franca et al. 1999, in preparation), while 
all our northern fields N1-3 have been completed to
a similar  depth using the INT Wide Field Camera (Verma et al. 1999, in
preparation).   Other optical bands allow object classification and
other more detailed investigations. Four
square degrees within our northern fields have been observed to a depth of $U\sim
22$ (Verma et al. 1999, in preparation). 
In June 1999, we observed the central 1.2 square degrees of S1 in $U$
and 3 square degrees in $I$ using the ESO Wide Field Imager
(H\'eraudeau   et al. 1999, in preparation) with these we expect to reach
$U\sim23, I\sim23$.
McMahon et al. (\verb+http://www.ast.cam.ac.uk/~rgm/int_sur/+)
have covered around 9 square degrees of N1 to 
$u\sim23.3,g\sim24.2,r\sim23.5,i\sim22.7, z\sim21.1$ and 2 square
degrees in N2 to similar depths in $g,r,i,z$
as part of the ING wide  field survey.   The $U$-band surveys will be
especially interesting as they will allow us to compute the $U$-band
luminosity density (and hence star formation rate) in the same volume
as we calculate the infrared luminosity density, providing a direct comparison
between obscured and unobscured star formation estimators.

\item {\bf Near Infrared:}
A substantial area has been surveyed in the near infrared. In the
$H$-band around 0.85 square degrees in N1 and N2 was surveyed using CIRSI on the
INT (Gonzalez-Solares et al. 1999, in preparation).  Approximately 0.5 square
degree has been surveyed in both N1 and N2 in $K^\prime$ using Omega
Prime on the Calar
Alto 3.5m (Rigopoulou et al. 1999, in preparation).  
The smaller, multiply repeated southern field S2 has been covered in
$K$ with SOFI on the  NTT (H\'eraudeau et al., 1999, in preparation).

\item {\bf Radio:}
21cm radio data at sub-mJy level will allow identification of some of
the most interesting objects which are expected to be  very
faint in the optical but would have detectable radio fluxes if they
obey the usual radio to far infrared
correlation.  These surveys will also allow an independent estimate of
the star formation rate within the same volume.  The southern field S1
is completely covered to a depth of 0.3 mJy \cite{Gruppioni et al. 1999a},
the 6 square degrees in the northern fields has been covered to a
depth of 0.2 mJy \cite{Ciliegi et al. 1999}.  A deeper survey in the south
has been conducted on the smaller, multiply repeated  field S2
(Gruppioni et al. 1999, in preparation).

\item{\bf X-ray:}
Almaini et al. have been awarded 150ks to do two deep {\em Chandra} pointings
one in N1 and one in N2.  La Franca et al. have also been awarded 200
ks  on  {\em BeppoSAX} to make 5 pointings covering around 2 square degrees
in S1.

\item{\bf Sub-mm:}
The UK SCUBA Survey consortium (Rowan-Robinson et al., independent of
{\em ELAIS}) are  performing part of their shallow (8mJy) 850\micron
survey in N1 and N2 and are aiming to cover 200 square arc minutes in each. 

\end{enumerate}

\begin{table*}

\begin{small}

\begin{tabular}{lccccccccccc}
Band       & 2-10keV &  $u,g,r,i,z$ & $R$   & $H$ & $K$ & 6.7 & 15  & 90 & 175  & 850  & 21cm  \\
Depth Units      & CGI       &   mag   & mag  & mag  & mag   & mJy & mJy & mJy & mJy & mJy  & mJy   \\
\\
\multicolumn{12}{c}{N1}\\
Area       & 0.07       &    9    & 3.1    &   0.5 & 0.4 &    & 2.6  & 2.6 &  2   & 0.05 &  1.54  \\
Depth      & $10^{-14}$
                      &   23.3,24.2,23.5,22.7,21.1    & 23  & 19.5 & 18.0 & 1   & 3   & 100 & 100 & 8 &  0.1-0.4 \\
\\
\multicolumn{12}{c}{N2}\\
Area       & 0.07       &    2    & 3.6    &  0.5   & 0.4&  2.7  & 2.7  & 2.7 &  1   & 0.05 &  1.54  \\
Depth      & $10^{-14}$     
                  &   22.5,24.2,23.5,22.7,21.1 & 23 & 19.5 & 18.0 & 1   & 3   & 100 & 100 & 8 &  0.1-0.4 \\

\\
\multicolumn{12}{c}{N3}\\

Area       &        &    1,1,2.3,1,0    & 2.3    &   1 &  &  1.32  & 0.9  & 1.76 &   &  &  1.14 \\
Depth      &     &   22.5,23,23,23,0    & 23  & 19.5 & & 1   & 3   & 100 &  &  &  0.1-0.4 \\
\\
\multicolumn{12}{c}{S1}\\

Area       & 2        & 1.2,0,4,3,0    & 4    &    &  &  1.8  & 4  & 4 &
&  &  4  \\
Depth      & $10^{-13}$
                      &   23,0,23.5,23,0    & 23.5  & &  & 1   & 3   & 100 &  &  &  0.24 \\

\end{tabular}
\end{small}
\caption{Multi-wavelength field surveys within the main {\em ELAIS} fields, the vast majority carried out as part 
of the {\em ELAIS} collaboration.  Areas are in square degrees. 
Some sub-fields within these go to greater depth.  The X-ray and
sub-mm surveys are yet to be completed.
}
\end{table*}

Additional multi-wavelength surveys of these fields are expected in
the near future.

\subsection{Photometry \& Spectroscopy}

We intend to obtain spectroscopic identifications for all (or the vast
majority) of  optical candidates for all {\em ELAIS} sources. This involves 
a two-pronged attack using
multi-object spectroscopy for the brightest objects and single object spectroscopy using 4m class telescopes on the fainter objects.
This will be principally to obtain the redshifts and thus
luminosity but also for classification and to assess
star formation rates. Some preliminary multi-fibre spectroscopy
has been carried out with FLAIR on the UK Schmidt Telescope. This been supplemented
by single object spectroscopy from the ESO/Danish 1.5m telescope to
provide spectroscopy on a complete sample of 90\micron\ selected sources
(Linden-V{\o}rnle et al. 1999, in preparation).  A further
100 sources have been identified 
spectroscopically in a largely weathered-out run on the  2dF in
September 1998  (Gruppioni et al. 1999, in preparation) and an additional night on the 2dF 
in August 1999 was also seriously hampered by weather (Oliver et al. 1999, in preparation).  40 spectra for fainter sources
have already been taken with EMMI on the NTT and EFOSC2 on the ESO
3.6m Telescope (La Franca et al. 1999, in preparation).

Until now the {\em ELAIS} northern fields have been only moderately surveyed 
spectroscopically. We have used the Calar Alto 2.2m telescope and the Calar Alto 
Faint Object Spectrograph (CAFOS) for the sources brighter than 17
(Gonzalez-Solares et al. 1999, in preparation) and the 
Calar Alto 3.5m telescope and the Multi Object Spectrograph (MOSCA) for the 
fainter sources (Surace et al., 1999, in preparation). 29 
{\em ELAIS} objects have been observed 
during the period May-July 1998 (of which 14 had $m>19$). 
These observations have been completed with 29 
field galaxies chosen in the same region for comparison purpose.
From these northern samples most sources show strong star-burst
signatures up to $z=0.5$ though two AGNs 
and one $z=1.2$ QSO have been detected, these samples will be discussed in a forthcoming paper.

A number of programmes have been instigated to obtain more specific  photometric and
spectroscopic data
of the infrared selected sources over a wider wavelength range.
Some examples are detailed below:

\begin{enumerate}
\item{}We have observed (H\'eraudeau, Kotilainen, Surace et al., in preparation) about 150 sources in pointing 
observations in the S1 field using IRAC2 on the ESO/MPG 2.2m telescope October 
1997, June 1998 and SOFI on the NTT October 1998
\item{} Near-infrared H+K band spectroscopy of a small subset of sources
with SOFI on the NTT (Alexander et al., in preparation).

\end{enumerate}

\section{Conclusions}

In this paper we have described the motivation behind ELAIS,
the largest non-serendipitous survey performed by {\em ISO\/}. Our
primary goals in conducting the survey were to determine the
relative importance and recent evolution of the dust--obscured mode of 
star formation in galaxies, and to constrain AGN unification models,
and we detailed above how these influenced our selection of survey
fields and observational parameters. 
The fields that have been covered by {\em ISO\/} are also being
extensively mapped from radio to X--ray wavelengths as part of a
concerted ground--based follow--up programme, whose multi--wavelength
coverage will make the {\em ELAIS\/} regions fertile ground for
undertaking future
astrophysical investigations extending well beyond our initial survey aims.

Subsequent papers in this series 
will discuss in detail the scientific results from the ELAIS ``Preliminary Analysis'' and ``Final Analysis''. The first of these papers will
include: discussions of the extra-galactic  counts from the
``Preliminary Analysis'' at 7 and 15 \micron
(Serjeant et al. 1999), and at  90$\mu$m (Efstathiou et al., 1999, in preparation); discussion of the stellar calibration and counts (Crockett et
al., in preparation); and a discussion of sources detected in the 
multiply--repeated areas (Oliver et al., in preparation).
Preliminary ELAIS data products were released through our WWW page 
(\verb+http://athena.ph.ic.ac.uk/+), which also contains further details
on the programme and the follow-up campaign.

\section*{Acknowledgments}

This paper is based on observations with {\em ISO}, an ESA project, with
instruments funded by ESA Member States (especially the PI countries:
France, Germany, the Netherlands and the United Kingdom) and with
participation of {\em ISAS} and {\em NASA}.

The {\em ISO-CAM} data presented in this paper was analysed using ``CIA'', 
a joint development by the ESA Astrophysics Division and the {\em ISO-CAM}    
Consortium. The {\em ISO-CAM} Consortium is led by the {\em ISO-CAM} PI, C. Cesarsky,  
Direction des Sciences de la Matiere, C.E.A., France.

     PIA is a joint development by the ESA Astrophysics Division 
     and the {\em ISOPHOT} Consortium.
     The {\em ISO-PHOT} Consortium is led by the Max-Planck-Institut fu\"{e}r 
     Astronomie (MPIA), Heidelberg, Germany.
     Contributing {\em ISO-PHOT} Consortium institutes to the PIA 
     development are: 
     DIAS (Dublin Institute for advanced studies, Ireland)
     MPIK (Max-Planck-Institut fu\"{e}r Kernphysik, Heidelberg, Germany),
     RAL (Rutherford Appleton Laboratory, Chilton, UK),
     AIP (Astronomisches Institut Potsdam, Germany), and MPIA.

This work in part was
supported by PPARC (grant 
number GR/K98728) and by the EC TMR Network programme
(FMRX-CT96-0068).

We would like to thank all the {\em ISO} staff at Vilspa both on the
science team and on the Instrument Development Teams  for their eternal
patience in dealing with the wide variety of problems that a large
programme like this presented.

\label{lastpage}

\appendix
\section{Log of the {\em ISO} Observations}\label{isolog}

In  Table A1 we present a list of all the observations
performed by {\em ISO} as part of the {\em ELAIS} raster observations.
We do not include the observations performed as part of the {\em ISO}
photometric follow-up of {\em ELAIS} sources which are available on
our WWW pages \verb+http://athena.ph.ic.ac.uk/+. Table \ref{tab:fail}
details those observations which have had some instrument or
telemetry problems as flagged at Vilspa or for which we have noted peculiarities.

\newpage

%

\begin{table*}
\caption{Log of all the ISO survey observations performed for  ELAIS,
excluding the pointed photometric observations. 
TDN is a unique identifier for each {\em ISO} observation and
monotonically increased throughout the mission.  OFFICIAL$\_$NAME is a
unique identifier for {\em ELAIS} observations constructed from the 
instrument name, the filter, the main field identifier and where
required a sub-field identifier and a multiplicity identifier, it does
not correspond to the name in the {\em ISO} archives.
}
\begin{tabular}{lllccccccll}
  TDN  &     OFFICIAL$\_$NAME &    RA (J2000)    Dec(J2000)&  ROLL &  M&
N &  DM &  DN &  TINT &  FILT &  STATUS \\ \hline 
11600721   & CAM$\_$LW3$\_$N2$\_$T$\_$I    & 16 35 45.00  +41 06 00.0   &
84   & 28   & 14   &  90   & 180   &  21   & LW3   & Observed \\
19201010   & PHT$\_$C90$\_$N1$\_$T$\_$I    & 16 10 01.20  +54 30 36.0   &
358   & 10   & 20   & 130   & 130   &  10   &  90   & Observed \\
19201091   & PHT$\_$C90$\_$N1$\_$T$\_$J    & 16 10 01.20  +54 30 36.0   &
358   & 10   & 20   & 130   & 130   &  32   &  90   & Observed \\
23200251   & CAM$\_$LW3$\_$S1$\_$1      & 00 30 25.40  -42 57 00.3   &  77
& 28   & 14   &  90   & 180   &  21   & LW3   & Observed \\
23200252   & PHT$\_$C90$\_$S1$\_$1A     & 00 30 14.90  -42 47 11.7   &  78
& 10   & 20   & 130   & 130   &  20   &  90   & Observed \\
23200289   & PHT$\_$C90$\_$S1$\_$1B     & 00 30 36.00  -43 06 48.8   &  78
& 10   & 20   & 130   & 130   &  20   &  90   & Observed \\
23200353   & CAM$\_$LW3$\_$S1$\_$2      & 00 31 08.20  -43 36 14.1   &  78
& 28   & 14   &  90   & 180   &  21   & LW3   & Observed \\
23200354   & PHT$\_$C90$\_$S1$\_$2A     & 00 30 57.40  -43 26 25.7   &  78
& 10   & 20   & 130   & 130   &  20   &  90   & Observed \\
23200392   & PHT$\_$C90$\_$S1$\_$2B     & 00 31 19.00  -43 46 02.5   &  78
& 10   & 20   & 130   & 130   &  20   &  90   & Observed \\
23300257   & CAM$\_$LW3$\_$S1$\_$4      & 00 33 59.40  -42 49 03.1   &  77
& 28   & 14   &  90   & 180   &  21   & LW3   & Observed \\
23300258   & PHT$\_$C90$\_$S1$\_$4A     & 00 33 48.30  -42 39 15.8   &  77
& 10   & 20   & 130   & 130   &  20   &  90   & Observed \\
23300294   & PHT$\_$C90$\_$S1$\_$4B     & 00 34 10.60  -42 58 50.4   &  78
& 10   & 20   & 130   & 130   &  20   &  90   & Observed \\
23300459   & CAM$\_$LW3$\_$S1$\_$5      & 00 34 44.40  -43 28 12.0   &  78
& 28   & 14   &  90   & 180   &  21   & LW3   & Observed \\
23300460   & PHT$\_$C90$\_$S1$\_$5A     & 00 34 33.10  -43 18 24.9   &  78
& 10   & 20   & 130   & 130   &  20   &  90   & Observed \\
23300495   & PHT$\_$C90$\_$S1$\_$5B     & 00 34 55.80  -43 37 59.1   &  78
& 10   & 20   & 130   & 130   &  20   &  90   & Observed \\
30200101   & CAM$\_$LW3$\_$N1$\_$1      & 16 15 01.00  +54 20 41.0   & 258
& 28   & 14   &  90   & 180   &  21   & LW3   & Observed \\
30200102   & PHT$\_$C90$\_$N1$\_$1A     & 16 15 16.70  +54 10 57.0   & 258
& 10   & 20   & 130   & 130   &  20   &  90   & Observed \\
30200113   & PHT$\_$C90$\_$N1$\_$1B     & 16 14 45.30  +54 30 24.9   & 258
& 10   & 20   & 130   & 130   &  20   &  90   & Observed \\
30400103   & CAM$\_$LW3$\_$N1$\_$2      & 16 13 57.10  +54 59 35.9   & 255
& 28   & 14   &  90   & 180   &  21   & LW3   &  Aborted \\
30400104   & PHT$\_$C90$\_$N1$\_$2A     & 16 14 13.30  +54 49 52.4   & 256
& 10   & 20   & 130   & 130   &  20   &  90   &   Failed \\
30400114   & PHT$\_$C90$\_$N1$\_$2B     & 16 13 40.90  +55 09 19.3   & 255
& 10   & 20   & 130   & 130   &  20   &  90   &  Aborted \\
30500105   & CAM$\_$LW3$\_$N1$\_$3      & 16 10 34.90  +54 11 12.7   & 254
& 28   & 14   &  90   & 180   &  21   & LW3   & Observed \\
30500106   & PHT$\_$C90$\_$N1$\_$3A     & 16 10 51.50  +54 01 30.9   & 254
& 10   & 20   & 130   & 130   &  20   &  90   & Observed \\
30500115   & PHT$\_$C90$\_$N1$\_$3B     & 16 10 18.10  +54 20 54.4   & 254
& 10   & 20   & 130   & 130   &  20   &  90   & Observed \\
30600107   & CAM$\_$LW3$\_$N1$\_$4      & 16 09 27.00  +54 49 58.7   & 253
& 28   & 14   &  90   & 180   &  21   & LW3   & Observed \\
30600108   & PHT$\_$C90$\_$N1$\_$4A     & 16 09 44.20  +54 40 17.4   & 253
& 10   & 20   & 130   & 130   &  20   &  90   & Observed \\
30600116   & PHT$\_$C90$\_$N1$\_$4B     & 16 09 09.70  +54 59 39.8   & 252
& 10   & 20   & 130   & 130   &  20   &  90   & Observed \\
30900111   & CAM$\_$LW3$\_$N1$\_$6      & 16 04 59.00  +54 39 44.3   & 249
& 28   & 14   &  90   & 180   &  21   & LW3   & Observed \\
30900112   & PHT$\_$C90$\_$N1$\_$6A     & 16 05 17.20  +54 30 05.5   & 249
& 10   & 20   & 130   & 130   &  20   &  90   & Observed \\
30900118   & PHT$\_$C90$\_$N1$\_$6B     & 16 04 40.70  +54 49 22.9   & 249
& 10   & 20   & 130   & 130   &  20   &  90   & Observed \\
31000109   & CAM$\_$LW3$\_$N1$\_$5      & 16 06 10.80  +54 01 08.0   & 248
& 28   & 14   &  90   & 180   &  21   & LW3   & Observed \\
31000117   & PHT$\_$C90$\_$N1$\_$5B     & 16 05 53.10  +54 10 47.3   & 248
& 10   & 20   & 130   & 130   &  20   &  90   & Observed \\
31000132   & PHT$\_$C90$\_$N1$\_$5A     & 16 06 28.40  +53 51 28.6   & 248
& 10   & 20   & 130   & 130   &  20   &  90   & Observed \\
40700479   & CAM$\_$LW3$\_$X6$\_$1      & 14 36 43.10  +15 44 13.0   & 124
& 12   &  6   &  90   & 180   &  21   & LW3   & Observed \\
40700480   & PHT$\_$C90$\_$X6$\_$1      & 14 36 43.10  +15 44 13.0   & 124
&  9   &  9   & 130   & 130   &  20   &  90   & Observed \\
40701983   & CAM$\_$LW3$\_$X4$\_$1      & 17 14 14.00  +50 15 23.6   & 163
& 12   &  6   &  90   & 180   &  21   & LW3   & Observed \\
40701984   & PHT$\_$C90$\_$X4$\_$1      & 17 14 14.00  +50 15 23.6   & 162
&  9   &  9   & 130   & 130   &  20   &  90   & Observed \\
40800464   & PHT$\_$C90$\_$S1$\_$7A     & 00 37 20.80  -42 30 55.3   & 251
& 10   & 20   & 130   & 130   &  20   &  90   & Observed \\
40800497   & PHT$\_$C90$\_$S1$\_$7B     & 00 37 44.20  -42 50 27.1   & 251
& 10   & 20   & 130   & 130   &  20   &  90   & Observed \\
40800663   & CAM$\_$LW3$\_$S1$\_$7      & 00 37 32.50  -42 40 41.2   & 251
& 28   & 14   &  90   & 180   &  21   & LW3   & Observed \\
40800765   & CAM$\_$LW3$\_$S1$\_$8      & 00 38 19.60  -43 19 44.5   & 251
& 28   & 14   &  90   & 180   &  21   & LW3   & Observed \\
41000856   & PHT$\_$C90$\_$S1$\_$3A     & 00 31 40.90  -44 05 38.9   & 254
& 10   & 20   & 130   & 130   &  20   &  90   & Observed \\
41000893   & PHT$\_$C90$\_$S1$\_$3B     & 00 32 03.00  -44 25 15.0   & 254
& 10   & 20   & 130   & 130   &  20   &  90   & Observed \\
41001062   & PHT$\_$C90$\_$S1$\_$6A     & 00 35 18.80  -43 57 33.0   & 253
& 10   & 20   & 130   & 130   &  20   &  90   & Observed \\
41001096   & PHT$\_$C90$\_$S1$\_$6B     & 00 35 42.00  -44 17 06.4   & 253
& 10   & 20   & 130   & 130   &  20   &  90   & Observed \\
41001161   & CAM$\_$LW3$\_$S1$\_$6      & 00 35 30.40  -44 07 19.8   & 253
& 28   & 14   &  90   & 180   &  21   & LW3   & Observed \\
41101867   & CAM$\_$LW3$\_$S1$\_$9      & 00 39 07.80  -43 58 46.6   & 253
& 28   & 14   &  90   & 180   &  21   & LW3   & Observed \\
41300766   & PHT$\_$C90$\_$S1$\_$8A     & 00 38 07.70  -43 09 58.8   & 255
& 10   & 20   & 130   & 130   &  20   &  90   &   Failed \\
41300798   & PHT$\_$C90$\_$S1$\_$8B     & 00 38 31.60  -43 29 30.2   & 255
& 10   & 20   & 130   & 130   &  32   &  90   &  Aborted \\
41300955   & CAM$\_$LW3$\_$S1$\_$3      & 00 31 51.90  -44 15 27.0   & 256
& 28   & 14   &  90   & 180   &  21   & LW3   & Observed \\
41301068   & PHT$\_$C90$\_$S1$\_$9A     & 00 38 55.70  -43 49 01.2   & 255
& 10   & 20   & 130   & 130   &  20   &  90   & Observed \\
41301099   & PHT$\_$C90$\_$S1$\_$9B     & 00 39 20.00  -44 08 31.8   & 255
& 10   & 20   & 130   & 130   &  20   &  90   & Observed \\
41502787   & CAM$\_$LW3$\_$X3$\_$1      & 00 22 48.00  -30 06 30.0   & 254
& 16   & 16   &  90   &  90   &  21   & LW3   & Observed \\
41502788   & PHT$\_$C90$\_$X3$\_$1      & 00 22 48.00  -30 06 30.0   & 254
& 12   & 12   & 130   & 130   &  20   &  90   & Observed \\
42500136   & PHT$\_$C90$\_$N3$\_$2A     & 14 25 49.90  +33 09 53.8   & 114
& 10   & 20   & 130   & 130   &  20   &  90   & Observed \\
42500146   & PHT$\_$C90$\_$N3$\_$2B     & 14 25 18.60  +32 51 00.7   & 114
& 10   & 20   & 130   & 130   &  20   &  90   & Observed \\
42500237   & CAM$\_$LW3$\_$N3$\_$3      & 14 29 38.30  +33 24 49.6   & 114
& 28   & 14   &  90   & 180   &  21   & LW3   & Observed \\
42700238   & PHT$\_$C90$\_$N3$\_$3A     & 14 29 54.50  +33 34 14.2   & 113
& 10   & 20   & 130   & 130   &  20   &  90   & Observed \\
42700247   & PHT$\_$C90$\_$N3$\_$3B     & 14 29 22.10  +33 15 24.9   & 113
& 10   & 20   & 130   & 130   &  20   &  90   & Observed \\
43800341   & CAM$\_$LW3$\_$N3$\_$5      & 14 32 38.20  +33 11 10.3   & 105
& 28   & 14   &  90   & 180   &  21   & LW3   & Observed \\
\end{tabular}
\end{table*}

\newpage
\begin{table*}
\contcaption{Log of all the ISO survey observations performed for  ELAIS,
excluding the pointed photometric observations.}
\begin{tabular}{lllccccccll}
  TDN  &     OFFICIAL$\_$NAME &    RA (J2000)    Dec(J2000)&  ROLL &  M&
N &  DM &  DN &  TINT &  FILT &  STATUS \\ \hline 
49900120   & PHT$\_$C90$\_$N2$\_$1A     & 16 32 33.00  +41 22 14.3   & 247
& 20   & 20   &  75   & 130   &  12   &  90   & Observed \\
49900222   & PHT$\_$C90$\_$N2$\_$1B     & 16 33 26.50  +41 04 51.7   & 247
& 20   & 20   &  75   & 130   &  12   &  90   & Observed \\
49900326   & PHT$\_$C90$\_$N2$\_$2B     & 16 35 10.50  +40 30 02.0   & 248
& 20   & 20   &  75   & 130   &  12   &  90   & Observed \\
50000124   & PHT$\_$C90$\_$N2$\_$2A     & 16 34 18.30  +40 47 27.6   & 247
& 20   & 20   &  75   & 130   &  12   &  90   & Observed \\
50000228   & PHT$\_$C90$\_$N2$\_$3A     & 16 35 39.40  +41 41 55.6   & 247
& 20   & 20   &  75   & 130   &  12   &  90   & Observed \\
50000330   & PHT$\_$C90$\_$N2$\_$3B     & 16 36 31.30  +41 24 27.7   & 247
& 20   & 20   &  75   & 130   &  12   &  90   & Observed \\
50000723   & CAM$\_$LW3$\_$N2$\_$3      & 16 36 05.50  +41 33 11.8   & 247
& 28   & 14   &  90   & 180   &  21   & LW3   & Observed \\
50100172   & PHT$\_$C90$\_$N2$\_$4B     & 16 38 14.20  +40 49 27.6   & 246
& 20   & 20   &  75   & 130   &  12   &  90   & Observed \\
50100273   & PHT$\_$C90$\_$N2$\_$5A     & 16 38 47.80  +42 01 18.0   & 246
& 20   & 20   &  75   & 130   &  12   &  90   & Observed \\
50100374   & PHT$\_$C90$\_$N2$\_$5B     & 16 39 39.70  +41 43 44.9   & 246
& 20   & 20   &  75   & 130   &  12   &  90   & Observed \\
50100727   & CAM$\_$LW3$\_$N2$\_$5      & 16 39 13.80  +41 52 31.6   & 246
& 28   & 14   &  90   & 180   &  21   & LW3   & Observed \\
50100871   & PHT$\_$C90$\_$N2$\_$4A     & 16 37 23.50  +41 06 58.3   & 246
& 20   & 20   &  75   & 130   &  12   &  90   & Observed \\
50200119   & CAM$\_$LW3$\_$N2$\_$1      & 16 32 59.80  +41 13 33.2   & 244
& 28   & 14   &  90   & 180   &  21   & LW3   & Observed \\
50200225   & CAM$\_$LW3$\_$N2$\_$4      & 16 37 48.90  +40 58 13.1   & 245
& 28   & 14   &  90   & 180   &  21   & LW3   & Observed \\
50200429   & CAM$\_$LW3$\_$N2$\_$6      & 16 40 55.50  +41 17 22.7   & 246
& 28   & 14   &  90   & 180   &  21   & LW3   & Observed \\
50200575   & PHT$\_$C90$\_$N2$\_$6A     & 16 40 30.10  +41 26 10.4   & 246
& 20   & 20   &  75   & 130   &  12   &  90   & Observed \\
51100131   & CAM$\_$LW3$\_$N2$\_$2      & 16 34 44.50  +40 38 45.0   & 236
& 28   & 14   &  90   & 180   &  21   & LW3   & Observed \\
51100234   & CAM$\_$LW2$\_$N2$\_$4      & 16 37 48.90  +40 58 13.1   & 236
& 28   & 14   &  90   & 180   &  21   & LW2   & Observed \\
51100736   & CAM$\_$LW2$\_$N2$\_$6      & 16 40 55.50  +41 17 22.7   & 237
& 28   & 14   &  90   & 180   &  21   & LW2   & Observed \\
51100835   & CAM$\_$LW2$\_$N2$\_$5      & 16 39 13.80  +41 52 31.6   & 236
& 28   & 14   &  90   & 180   &  21   & LW2   & Observed \\
51200131   & CAM$\_$LW2$\_$N2$\_$1      & 16 32 59.80  +41 13 33.2   & 234
& 28   & 14   &  90   & 180   &  21   & LW2   & Observed \\
51200232   & CAM$\_$LW2$\_$N2$\_$2      & 16 34 44.50  +40 38 45.0   & 235
& 28   & 14   &  90   & 180   &  21   & LW2   & Observed \\
51200433   & CAM$\_$LW2$\_$N2$\_$3      & 16 36 05.50  +41 33 11.8   & 235
& 28   & 14   &  90   & 180   &  21   & LW2   & Observed \\
51200576   & PHT$\_$C90$\_$N2$\_$6B     & 16 41 20.80  +41 08 34.6   & 236
& 20   & 20   &  75   & 130   &  12   &  90   & Observed \\
54502485   & CAM$\_$LW3$\_$X1$\_$1      & 01 13 12.80  -45 14 06.7   &  33
& 16   &  8   &  90   & 180   &  21   & LW3   & Observed \\
54502486   & PHT$\_$C90$\_$X1$\_$1      & 01 13 12.80  -45 14 06.7   &  33
& 12   & 12   & 130   & 130   &  20   &  90   & Observed \\
59800143   & CAM$\_$LW2$\_$S1$\_$5      & 00 34 44.40  -43 28 12.0   &  76
& 28   & 14   &  90   & 180   &  21   & LW2   & Observed \\
59800244   & CAM$\_$LW2$\_$S1$\_$6      & 00 35 30.40  -44 07 19.8   &  77
& 28   & 14   &  90   & 180   &  21   & LW2   & Observed \\
59800745   & CAM$\_$LW2$\_$S1$\_$8      & 00 38 19.60  -43 19 44.5   &  77
& 28   & 14   &  90   & 180   &  21   & LW2   & Observed \\
59800846   & CAM$\_$LW2$\_$S1$\_$9      & 00 39 07.80  -43 58 46.6   &  77
& 28   & 14   &  90   & 180   &  21   & LW2   & Observed \\
61300341   & CAM$\_$LW2$\_$N3$\_$5      & 14 32 38.20  +33 11 10.3   & 291
& 28   & 14   &  90   & 180   &  21   & LW2   & Observed \\
61600642   & PHT$\_$C90$\_$N3$\_$5A     & 14 32 54.70  +33 20 33.4   & 289
& 20   & 20   &  75   & 130   &  12   &  90   & Observed \\
61600649   & PHT$\_$C90$\_$N3$\_$5B     & 14 32 21.70  +33 01 47.0   & 289
& 20   & 20   &  75   & 130   &  12   &  90   & Observed \\
61800277   & CAM$\_$LW3$\_$X2$\_$1      & 13 34 36.00  +37 54 36.0   & 280
& 16   &  8   &  90   & 180   &  21   & LW3   & Observed \\
61800278   & PHT$\_$C90$\_$X2$\_$1      & 13 34 36.00  +37 54 36.0   & 279
& 12   & 12   & 130   & 130   &  20   &  90   & Observed \\
62300324   & CAM$\_$LW2$\_$N3$\_$6D     & 14 32 01.00  +32 20 47.0   & 285
& 14   &  7   &  90   & 180   &  21   & LW2   & Observed \\
62300422   & CAM$\_$LW2$\_$N3$\_$6B     & 14 30 32.00  +32 27 37.0   & 284
& 14   &  7   &  90   & 180   &  21   & LW2   & Observed \\
63500448   & PHT$\_$C90$\_$N3$\_$6D     & 14 32 01.00  +32 20 47.0   & 277
& 10   & 20   & 130   &  65   &  12   &  90   & Observed \\
63500539   & PHT$\_$C90$\_$N3$\_$4C     & 14 29 34.80  +32 53 13.0   & 276
& 10   & 20   & 130   &  65   &  12   &  90   & Observed \\
63500628   & PHT$\_$C90$\_$N3$\_$1D     & 14 27 06.80  +33 25 28.0   & 275
& 10   & 20   & 130   &  65   &  12   &  90   & Observed \\
63500726   & PHT$\_$C90$\_$N3$\_$1B     & 14 25 36.30  +33 32 04.0   & 275
& 10   & 20   & 130   &  65   &  12   &  90   & Observed \\
63500825   & PHT$\_$C90$\_$N3$\_$1A     & 14 26 07.90  +33 50 57.0   & 275
& 10   & 20   & 130   &  65   &  12   &  90   & Observed \\
63501041   & PHT$\_$C90$\_$N3$\_$5C$\_$I   & 14 32 09.70  +33 24 00.0   &
276   & 10   & 20   & 130   &  65   &  12   &  90   & Observed \\
63501142   & PHT$\_$C90$\_$N3$\_$5D$\_$I   & 14 31 36.90  +33 05 13.0   &
276   & 10   & 20   & 130   &  65   &  12   &  90   & Observed \\
63800106   & CAM$\_$LW2$\_$N3$\_$2B     & 14 24 33.60  +32 54 17.0   & 273
& 14   &  7   &  90   & 180   &  21   & LW2   & Observed \\
63800205   & CAM$\_$LW2$\_$N3$\_$2A     & 14 25 04.80  +33 13 11.0   & 273
& 14   &  7   &  90   & 180   &  21   & LW2   & Observed \\
63800302   & CAM$\_$LW2$\_$N3$\_$1B     & 14 25 36.30  +33 32 04.0   & 273
& 14   &  7   &  90   & 180   &  21   & LW2   & Observed \\
63800401   & CAM$\_$LW2$\_$N3$\_$1A     & 14 26 07.90  +33 50 57.0   & 273
& 14   &  7   &  90   & 180   &  21   & LW2   & Observed \\
63800504   & CAM$\_$LW2$\_$N3$\_$1D     & 14 27 06.80  +33 25 28.0   & 272
& 14   &  7   &  90   & 180   &  21   & LW2   & Observed \\
63800603   & CAM$\_$LW2$\_$N3$\_$1C     & 14 27 38.70  +33 44 19.0   & 272
& 14   &  7   &  90   & 180   &  21   & LW2   & Observed \\
63800745   & PHT$\_$C90$\_$N3$\_$6A     & 14 31 04.40  +32 46 25.0   & 273
& 10   & 20   & 130   &  65   &  12   &  90   &   Failed \\
67200103   & CAM$\_$LW3$\_$N1$\_$2$\_$I    & 16 13 57.10  +54 59 35.9   &
254   & 28   & 14   &  90   & 180   &  21   & LW3   & Observed \\
67200114   & PHT$\_$C90$\_$N1$\_$2C     & 16 12 49.40  +54 57 15.1   & 254
& 20   & 20   & 130   &  65   &  20   &  90   & Observed \\
67500104   & PHT$\_$C90$\_$N1$\_$1C$\_$I   & 16 13 54.30  +54 18 22.3   &
251   & 20   & 20   & 130   &  65   &  12   &  90   & Observed \\
75200310   & PHT$\_$C90$\_$N1$\_$T$\_$K    & 16 11 09.60  +54 31 52.7   &
171   & 10   & 20   & 130   & 130   &  12   &  90   & Observed \\
75200411   & PHT$\_$C90$\_$N1$\_$T$\_$L    & 16 08 52.90  +54 29 16.9   &
170   & 10   & 20   & 130   & 130   &  12   &  90   & Observed \\
75200512   & PHT$\_$C90$\_$N1$\_$T$\_$M    & 16 09 52.10  +54 40 30.9   &
170   & 20   & 10   & 130   & 130   &  12   &  90   & Observed \\
75200613   & PHT$\_$C90$\_$N1$\_$T$\_$F    & 16 10 10.10  +54 20 41.1   &
171   & 20   & 10   & 130   & 130   &  12   &  90   & Observed \\
75901219   & PHT$\_$C90$\_$S1$\_$5$\_$L    & 00 34 44.40  -43 28 12.0   &
242   & 14   & 14   & 130   & 130   &  12   &  90   & Observed \\
76500218   & PHT$\_$C90$\_$S1$\_$5$\_$K    & 00 34 44.40  -43 28 12.0   &
245   & 14   & 14   & 130   & 130   &  12   &  90   & Observed \\
\end{tabular}
\end{table*}

\newpage
\begin{table*}

\contcaption{Log of all the ISO survey observations performed for  ELAIS,
excluding the pointed photometric observations.}
\begin{tabular}{lllccccccll}
  TDN  &     OFFICIAL$\_$NAME &    RA (J2000)    Dec(J2000)&  ROLL &  M&
N &  DM &  DN &  TINT &  FILT &  STATUS \\ \hline 
77001315   & PHT$\_$C90$\_$N2$\_$T$\_$L    & 16 34 58.80  +41 01 05.1   &
157   & 10   & 20   & 130   & 130   &  12   &  90   & Observed \\
77100214   & PHT$\_$C90$\_$N2$\_$T$\_$K    & 16 36 31.30  +41 10 53.8   &
156   & 10   & 20   & 130   & 130   &  12   &  90   & Observed \\
77400316   & PHT$\_$C90$\_$N2$\_$T$\_$M    & 16 35 18.90  +41 14 42.7   &
153   & 20   & 10   & 130   & 130   &  12   &  90   & Observed \\
77400417   & PHT$\_$C90$\_$N2$\_$T$\_$F    & 16 36 11.00  +40 57 17.0   &
153   & 20   & 10   & 130   & 130   &  12   &  90   & Observed \\
77500166   & PHT$\_$C90$\_$S1$\_$8A     & 00 38 07.70  -43 09 58.8   & 252
& 20   & 20   &  65   & 130   &  12   &  90   & Observed \\
77500207   & CAM$\_$LW3$\_$S1$\_$5$\_$K    & 00 34 44.40  -43 28 12.0   &
252   & 28   & 14   &  90   & 180   &  21   & LW3   & Observed \\
77800367   & CAM$\_$LW3$\_$N1$\_$U$\_$J    & 16 11 00.40  +54 13 25.4   &
145   &  8   &  4   &  90   & 180   &  21   & LW3   & Observed \\
77800368   & CAM$\_$LW3$\_$N1$\_$U$\_$K    & 16 11 00.40  +54 13 25.4   &
145   &  8   &  4   &  90   & 180   &  21   & LW3   & Observed \\
77800369   & CAM$\_$LW3$\_$N1$\_$U$\_$L    & 16 11 00.50  +54 13 31.3   &
145   &  8   &  4   &  90   & 180   &  21   & LW3   & Observed \\
77800370   & CAM$\_$LW3$\_$N1$\_$U$\_$M    & 16 11 00.50  +54 13 31.3   &
145   &  8   &  4   &  90   & 180   &  21   & LW3   & Observed \\
77800371   & CAM$\_$LW3$\_$N1$\_$U$\_$N    & 16 11 00.90  +54 13 21.7   &
145   &  8   &  4   &  90   & 180   &  21   & LW3   & Observed \\
77800372   & CAM$\_$LW3$\_$N1$\_$U$\_$O    & 16 11 00.90  +54 13 21.7   &
145   &  8   &  4   &  90   & 180   &  21   & LW3   & Observed \\
77900101   & CAM$\_$LW3$\_$N2$\_$T$\_$J    & 16 35 45.00  +41 06 00.0   &
148   & 28   & 14   &  90   & 180   &  21   & LW3   & Observed \\
77900202   & CAM$\_$LW2$\_$N2$\_$T$\_$J    & 16 35 45.00  +41 06 00.0   &
148   & 28   & 14   &  90   & 180   &  21   & LW2   & Observed \\
78500120   & PHT$\_$C160$\_$N2$\_$U$\_$I   & 16 35 50.00  +41 32 33.1   &
142   & 13   & 13   &  96   &  96   &  16   & 160   & Observed \\
78500221   & PHT$\_$C160$\_$N2$\_$U$\_$J   & 16 35 44.10  +41 31 29.0   &
142   & 13   & 13   &  96   &  96   &  16   & 160   & Observed \\
78500322   & PHT$\_$C160$\_$N2$\_$V$\_$I   & 16 34 39.70  +41 19 42.7   &
141   & 13   & 13   &  96   &  96   &  16   & 160   & Observed \\
78502406   & CAM$\_$LW3$\_$S1$\_$5$\_$J    & 00 34 44.40  -43 28 12.0   &
261   & 28   & 14   &  90   & 180   &  21   & LW3   & Observed \\
78600108   & CAM$\_$LW2$\_$S1$\_$5$\_$J    & 00 34 44.40  -43 28 12.0   &
261   & 28   & 14   &  90   & 180   &  21   & LW2   & Observed \\
78700123   & PHT$\_$C160$\_$N2$\_$V$\_$J   & 16 34 33.80  +41 18 38.6   &
139   & 13   & 13   &  96   &  96   &  16   & 160   & Observed \\
78700224   & PHT$\_$C160$\_$N2$\_$W$\_$I   & 16 33 29.80  +41 06 49.7   &
139   & 13   & 13   &  96   &  96   &  16   & 160   & Observed \\
78700325   & PHT$\_$C160$\_$N2$\_$W$\_$J   & 16 33 24.00  +41 05 45.6   &
139   & 13   & 13   &  96   &  96   &  16   & 160   & Observed \\
78800126   & PHT$\_$C160$\_$N2$\_$X$\_$I   & 16 36 58.20  +41 19 19.9   &
139   & 13   & 13   &  96   &  96   &  16   & 160   & Observed \\
78800227   & PHT$\_$C160$\_$N2$\_$X$\_$J   & 16 36 52.40  +41 18 15.8   &
139   & 13   & 13   &  96   &  96   &  16   & 160   & Observed \\
78800328   & PHT$\_$C160$\_$N2$\_$Y$\_$I   & 16 35 47.90  +41 06 32.0   &
139   & 13   & 13   &  96   &  96   &  16   & 160   & Observed \\
79400232   & PHT$\_$C160$\_$N2$\_$E$\_$I   & 16 38 06.00  +41 06 04.1   &
133   & 13   & 13   &  96   &  96   &  16   & 160   & Observed \\
79400333   & PHT$\_$C160$\_$N2$\_$E$\_$J   & 16 38 00.20  +41 05 00.0   &
133   & 13   & 13   &  96   &  96   &  16   & 160   & Observed \\
79400434   & PHT$\_$C160$\_$N2$\_$F$\_$I   & 16 36 55.70  +40 53 18.9   &
133   & 13   & 13   &  96   &  96   &  16   & 160   & Observed \\
79500329   & PHT$\_$C160$\_$N2$\_$Y$\_$J   & 16 35 42.10  +41 05 27.9   &
132   & 13   & 13   &  96   &  96   &  16   & 160   & Observed \\
79500430   & PHT$\_$C160$\_$N2$\_$Z$\_$I   & 16 34 38.10  +40 53 41.6   &
132   & 13   & 13   &  96   &  96   &  16   & 160   & Observed \\
79500531   & PHT$\_$C160$\_$N2$\_$Z$\_$J   & 16 34 32.30  +40 52 37.5   &
132   & 13   & 13   &  96   &  96   &  16   & 160   & Observed \\
79600173   & CAM$\_$LW3$\_$N1$\_$U$\_$P    & 16 11 00.40  +54 13 25.4   &
127   &  8   &  4   &  90   & 180   &  21   & LW3   & Observed \\
79600174   & CAM$\_$LW3$\_$N1$\_$U$\_$Q    & 16 11 00.40  +54 13 25.4   &
127   &  8   &  4   &  90   & 180   &  21   & LW3   & Observed \\
79600175   & CAM$\_$LW3$\_$N1$\_$U$\_$R    & 16 10 59.70  +54 13 23.1   &
127   &  8   &  4   &  90   & 180   &  21   & LW3   & Observed \\
79600176   & CAM$\_$LW3$\_$N1$\_$U$\_$I    & 16 10 59.70  +54 13 23.1   &
127   &  8   &  4   &  90   & 180   &  21   & LW3   & Observed \\
79800135   & PHT$\_$C160$\_$N2$\_$F$\_$J   & 16 36 49.90  +40 52 14.8   &
129   & 13   & 13   &  96   &  96   &  16   & 160   & Observed \\
79800236   & PHT$\_$C160$\_$N2$\_$G$\_$I   & 16 35 45.90  +40 40 30.9   &
129   & 13   & 13   &  96   &  96   &  16   & 160   & Observed \\
79800337   & PHT$\_$C160$\_$N2$\_$G$\_$J   & 16 35 40.10  +40 39 26.9   &
129   & 13   & 13   &  96   &  96   &  16   & 160   & Observed \\
80600181   & CAM$\_$LW3$\_$X5$\_$1      & 02 14 17.20  -11 58 46.2   & 254
& 12   & 12   &  90   &  90   &  21   & LW3   & Observed \\
86800640   & PHT$\_$C90$\_$S2$\_$1$\_$J    & 05 02 25.70  -30 36 24.0   &
287   & 18   &  9   &  65   & 130   &  12   &  90   & Observed \\
86901244   & PHT$\_$C90$\_$S2$\_$1$\_$L    & 05 02 22.10  -30 36 21.9   &
288   &  9   &  9   & 130   & 130   &  20   &  90   & Observed \\
86901341   & CAM$\_$LW3$\_$S2$\_$1$\_$J    & 05 02 24.30  -30 36 04.7   &
288   & 14   &  7   &  90   & 180   &  21   & LW3   & Observed \\
86901445   & CAM$\_$LW3$\_$S2$\_$1$\_$L    & 05 02 23.60  -30 36 04.3   &
288   & 14   &  7   &  90   & 180   &  21   & LW3   & Observed \\
86901539   & CAM$\_$LW3$\_$S2$\_$1$\_$I    & 05 02 24.00  -30 36 00.0   &
288   & 14   &  7   &  90   & 180   &  21   & LW3   & Observed \\
87001338   & PHT$\_$C90$\_$S2$\_$1$\_$I    & 05 02 24.00  -30 36 00.0   &
289   &  9   & 18   & 130   &  65   &  12   &  90   & Observed \\
87001442   & PHT$\_$C90$\_$S2$\_$1$\_$K    & 05 02 25.90  -30 35 38.1   &
289   & 18   &  9   &  65   & 130   &  12   &  90   & Observed \\
87001743   & CAM$\_$LW3$\_$S2$\_$1$\_$K    & 05 02 24.40  -30 35 55.7   &
289   & 14   &  7   &  90   & 180   &  21   & LW3   & Observed \\
87001846   & PHT$\_$C90$\_$S2$\_$1$\_$M    & 05 02 22.30  -30 35 36.0   &
289   &  9   &  9   & 130   & 130   &  20   &  90   & Observed \\
87001909   & PHT$\_$C160$\_$S2$\_$1$\_$M   & 05 02 22.30  -30 35 36.0   &
289   & 12   & 12   &  96   &  96   &  32   & 160   & Observed \\
87500403   & PHT$\_$C160$\_$S2$\_$1$\_$K   & 05 02 25.90  -30 35 38.1   &
293   &  6   & 12   & 180   &  96   &  32   & 160   & Observed \\
87500705   & PHT$\_$C160$\_$S2$\_$1$\_$L   & 05 02 22.10  -30 36 21.9   &
293   &  6   & 12   & 180   &  96   &  32   & 160   & Observed \\
87500808   & CAM$\_$LW2$\_$S2$\_$1$\_$M    & 05 02 23.70  -30 35 55.3   &
293   & 14   &  7   &  90   & 180   &  21   & LW2   & Observed \\
\end{tabular}
\end{table*}

\begin{table}

\caption{List of observations with telemetry or similar instrument problems
}\label{tab:fail}

\begin{tabular}{lll}
 TDN    & Name & Description \\
19201091 & PHT$\_$C90$\_$N1$\_$T$\_$J   & Very high glitch rate due to position in orbit\\
23300257 & CAM$\_$LW3$\_$S1$\_$4  & Telemetry drops caused some science data to be lost\\
30400104 &  PHT$\_$C90$\_$N1$\_$2A  & failed due to telemetry drops \\
41001062 &  PHT$\_$C90$\_$S1$\_$6A & Vilspa flagged as ``Unknown quality''\\
41001161 &  CAM$\_$LW3$\_$S1$\_$6 & Vilspa flagged as ``Unknown quality''\\
41300766 &   PHT$\_$C90$\_$S1$\_$8A  & failed due to telemetry drops \\
63800745 &  PHT$\_$C90$\_$N3$\_$6A & failed due to instrument problems\\
77400417 &  PHT$\_$C90$\_$N2$\_$T$\_$F& Vilspa flagged as ``Unknown quality''\\
87500403 & PHT$\_$C160$\_$S2$\_$1$\_$K  & Instrument problems (warm up)\\
87500705 &  PHT$\_$C160$\_$S2$\_$1$\_$L  & Instrument problems (warm up)\\
\end{tabular}

\end{table}



\end{document}